\documentclass{mn2e}
\usepackage{epsfig,natbib2,natbibmnfix}
\usepackage{amsmath}
\usepackage{amssymb}
\usepackage{mathrsfs}
\usepackage{subfigure}
\usepackage{url}
\usepackage{xcolor}
\usepackage{wrapfig}

\newcommand{\be}{\begin{equation}}
\newcommand{\ee}{\end{equation}}
\newcommand{\apj}{ApJ}
\newcommand{\apjs}{ApJS}
\newcommand{\mnras}{MNRAS}
\newcommand{\aap}{A\&A}
\newcommand{\araa}{ARA\&A}
\newcommand{\apjl}{ApJL}
\newcommand{\aj}{AJ}

\newcommand{\aapr}{AAPR}

\def\ltsima{$\; \buildrel < \over \sim \;$}
\def\simlt{\lower.5ex\hbox{\ltsima}}
\def\gtsima{$\; \buildrel > \over \sim \;$}
\def\simgt{\lower.5ex\hbox{\gtsima}}

\newcommand\Mdot{\dot{M}}
\def\del#1{{}}

\def\msun{{\,{\rm M}_\odot}}

\def\Gizmo{{\sc Gizmo }}

\title[Positive feedback at the disc-halo interface]
{Positive feedback at the disc-halo interface}

\author[Alexander Hobbs \& Robert Feldmann]
       {\parbox{18cm}{Alexander Hobbs$^{1}$ \& Robert Feldmann$^{1}$}\vspace{0.3cm}\\
\noindent $^{1}$Institute for Computational Science, University of Zurich, Winterthurerstrasse 190, CH-8057 Zurich, Switzerland}

\begin{document}

\maketitle

\begin{abstract}
The flat star formation (SF) history of the Milky Way (MW) requires gas in the Galactic disc to be replenished, most likely from a reservoir outside the Galaxy. Such a replenishment may be achieved by a form of `positive' feedback, whereby SF feedback creates a Galactic fountain cycle that collects and cools additional gas from the hot halo surrounding the Galaxy. In this paper we present a model of this process for the MW. A section of the Galactic disc is allowed to form stars which subsequently explode as supernovae (SNe) and send gas out into the hot halo. The gas that is sent out is colder than the hot halo gas and, as it mixes, the halo gas is cooled, providing fuel for further SF as the mixture falls back onto the Galactic disc. We find that this process can be sufficient to maintain a roughly-constant cold gas mass in the MW over at least $3$ Gyr. Our results further suggest that there is a positive feedback trend  whereby increasing SF leads to an increase in the cold gas budget at average SF rates below $0.5 \msun$yr$^{-1}$ but above which becomes negative, where further increasing the SFR causes the cold gas budget to decrease. We have constructed an analytical model for this that reproduces the data well and could have profound implications for galaxy evolution in feedback-dominated regimes.
\end{abstract}

\renewcommand{\thefootnote}{\fnsymbol{footnote}}
\footnotetext[1]{E-mail: {\tt alexander.p.hobbs@gmail.com}}

\section{Introduction}

The presence of a hot, gaseous halo around the Milky Way (MW) and other, late-type spiral galaxies \citep[for the lines of evidence for this see, e.g.,][]{KacprzakEtal2008, GaenslerEtal2008, NicastroEtal2008, MastropietroEtal2005, GrcevichPutman2009, BogdanEtal2013, MillerBregman2015, AndersonEtal2016} has opened up a realm of scientific inquiry into the relationship between the galactic disc and the hot, low-density gas that surrounds it. A link between this halo gas and the star-forming gas in the galaxy exists through a process known as a `galactic fountain', whereby feedback from supernovae (SNe) in the disc causes gas to be ejected out of the plane and into the disc-halo interface which extends to $\sim 2-3$ kpc. The ejected gas may later fall back onto the disc due to gravity \citep{ShapiroField1976, Bregman1980, MelioliEtal2008a} and provide fresh fuel for star formation, allowing the cycle to repeat again.

This cycle of ejection and re-accretion has been touted as a possible explanation for the near-constant SF rate (SFR) in our own Galaxy over the last $\sim 8$\,Gyr since $z \sim 1$ \citep[e.g.,][]{Rocha-PintoEtal2000}. This statistic stands in contrast to the SFR of the wider Universe, which shows a dramatic decline since redshift $z \sim 2-3$ \citep[e.g.,][]{LillyEtal1996, MadauEtal1998, HippeleinEtal2003}. A near-constant rate implies that the gas used up in star formation (SF) must be constantly replenished, over nearly a Hubble time. Accretion from the larger cosmological environment can likely only play a part until $z \simeq 1$ \citep{KeresEtal2005,  FernandezEtal2012, NelsonEtal2013, WoodsEtal2014}, and mergers with other galaxies likely only account for a $\sim$ tenth of what is required \citep[see][for a review]{SancisiEtal2008}. Recycling of gas by stellar winds may be able to account for closer to half of what is required at low redshift ($z < 0.5$) but it is not clear whether this gas would be returned to the disc or would end up in the halo \citep{LeitnerKravtsov2011}. Wind recycling moreover suffers from a law of diminishing returns and would be unable to provide the required low-metallicity gas to remedy the G-dwarf problem \citep{Schmidt1963, WortheyEtal1996} and related K- and M-dwarf problems \citep{CasusoBeckman2004, WoolfWest2012}. There must therefore be an alternative mechanism, not yet established in the scientific community, and it likely relates to the galactic fountain process -- a form of `\emph{positive feedback}' whereby the action of supernovae (SNe), rather than to get rid of gas, is to bring more gas into the Galaxy that can contribute to the star-forming gas budget.

Observations of the disc-halo interface have begun to provide a wealth of information in recent years. Surveys have unmasked the presence of extra-planar cold atomic (H {\scriptsize I}) gas in the vicinity of the MW \citep[for a review see, e.g.,][]{PutmanEtal2012}. Of particular interest are those extra-planar structures termed `intermediate-velocity clouds' (IVCs), that possess radial velocities with respect to the local standard of rest (LSR) of $40$ km s$^{-1} \leq \vert v_{\rm LSR} \vert \leq 90$ km s$^{-1}$. The IVCs inhabit a region in the halo within $\sim 2$ kpc from the disc plane \citep{RohserEtal2016}. Their properties, including their near-solar metallicity, suggest that they may be connected to a fountain process occurring in our own Galaxy. A class of IVCs of particular interest are those that contain significant amounts of molecular gas, allowing them to be detected in carbon monoxide (CO) emission \citep{MagnaniSmith2010}. These are known as molecular IVCs (MIVCs), and they appear to be quite numerous. \cite{RohserEtal2016} employ a method using far-infrared (FIR) excess emission to probe small angular scales, and find 239 MIVC candidates over both Galactic hemispheres with both negative and positive radial velocities, within a latitude limit of 20$^{\circ}$. This distribution of MIVC candidates is in good agreement with the rotational model of \cite{MarascoFraternali2011}, which already lends support to the picture of IVCs as originating from galactic fountain gas. Extrapolating to the entire MW, \cite{RohserEtal2016} find that the infalling IVC and MIVC gas may account for the main fraction of the gas inflow onto the disc. MIVCs may therefore be a late stage in the Galactic fountain process.


Recently, the presence of molecular gas in galaxies has been found to correlate with the mass, and density, of the hot `atmosphere' surrounding giant ellipticals and early spiral galaxies \citep{BabykEtal2018}. The current interpretation of this finding is that cooling from the hot atmosphere is the origin of the molecular gas. In particular, it is a cooling time of $< 1$ Gyr within $10$ kpc that allows for the formation of molecular gas; this is the `cooling time threshold' \citep{CavagnoloEtal2008, RaffertyEtal2008, HoganEtal2017b} which has been observed to be a reasonably sharp cut-off in the presence of molecular gas in early-types and brightest cluster galaxies (BCGs). In early-types, it is only possible to get molecular gas if you are below this cooling time threshold -- but in spiral galaxies molecular gas is present regardless \citep[e.g.,][]{Wilson2013}. This points to the conclusion that in early-types it is the simple condensation of the hot corona that gives rise to molecular gas, but in spiral galaxies such as the MW there is an additional mechanism occurring that does the job. Recent observations of the massive outflow in NGC6240 have begun to probe the origins of the molecular gas, finding that a significant fraction of it is entrained in the outflow itself rather than residing in the galaxy \citep{CiconeEtal2018}.




On the theoretical side, simulations have typically seen stellar feedback in the galactic disc as a form of negative feedback, by both heating gas and by ejecting gas out of the galaxy, thereby leaving less `cold' gas available to form stars. Indeed, in low-mass halos, stellar feedback is invoked as resolving the discrepancy between the halo mass function and the galaxy mass function \citep[e.g.,][]{PuchweinSpringel2013}. However, for halos more massive than a few times $10^{10} \msun$ the picture of stellar feedback as playing a negative role in the gas supply of the galaxy has recently been challenged \citep{FraternaliBinney2008, MarinacciEtal2011, HobbsEtal2013, HobbsEtal2015, ArmillottaEtal2016}.

The problem with invoking a hot halo as the source of the continued SF is that such a feature is extremely slow to cool and stable to thermal instabilities \citep{JoungEtal2012}. Early simulations attempting to model galaxies with hot halos found erroneous cooling, usually taking the form of a sudden condensation into cold gas clumps, that is now known to be due to numerical instabilities in the ‘classic’ smoothed particle hydrodynamics (SPH) technique \citep{HobbsEtal2013, NelsonEtal2013} and does not feature in simulations conducted with newer techniques that do not suffer from these (particular) numerical errors. Of course, without cooling, the gas is hydrostatically supported and cannot add to the cold gas budget of the disc at the required rate. If gas is to cool from the hot halo, therefore, some mechanism is needed that can excite a thermal instability. One such mechanism was proposed by \cite{MarinacciEtal2011}, in which a galactic fountain seeds metal-rich gas into the metal-poor hot halos, giving rise to a thermal instability that causes cold gas to rain down onto the disc in the form of $\sim 10^5 \msun$ clouds \citep[see also][]{FraternaliBinney2008}. This model provides an excellent fit to both the velocity and spatial distributions of warm IVC-like HI gas in the MW \citep{MarascoEtal2012}. 2D and 3D simulations of individual cold (T $\approx 10^4$ K) clouds moving through a hot medium were presented by \cite{MarinacciEtal2011} and subsequent follow-up papers \citep[e.g.,][]{ArmillottaEtal2016}, where the turbulent mixing of the cloud with the hot gas and subsequent cooling caused the mass of cold gas to increase by $\sim 30$ \% over a period of $60$ Myr. Similar work by \cite{GronkeOh2018}, investigating the formation of an extended `cooling tail' that continues to draw in and cool hot gas through thermal pressure gradients, found an increase in dense gas mass in the cloud of an order of magnitude over $\approx 20$ cloud-crushing times. 

Such clouds are presumed to be the ejecta from the disc via SNe feedback, and travel through the hot halo, entraining and cooling halo gas before returning to the disc and delivering this gas as fuel for further SF. The key concept here is that the cooling of gas from the halo is therefore dependent on the presence of star-forming gas deeper inside the potential well. The \cite{MarinacciEtal2011, ArmillottaEtal2016} papers argue that this model is sensitive to galactic environment, i.e., in halos above $\sim 10^{13} \msun$ the process is ineffective. This implies a quenching of SF and an inability to cool the hot corona going from late-type to early-type systems. 

Such a mechanism constitutes a positive feedback mode of gas accretion onto the disc -- where, rather than hindering SF and reducing the cold gas budget, feedback from SNe can do the opposite, acting as a means to `refrigerate' the hot halo and encourage it to cool and accrete onto the galactic disc.

Many simulations of late-time star formation in galaxies -- particularly those using early versions of SPH -- were plagued by numerical artifacts that prevented a clear picture of how galactic cold gas accretion proceeds. There has been much discussion, debate, and development in this area in the past decade, however, and the latest codes are substantially more reliable \citep{Hopkins2015}, allowing for useful experimentation to be carried out with galaxy formation models. Such models now possess improvements in physical modelling techniques, such as second-order accurate radiative cooling, chemical evolution, improved SF and feedback with metallicity evolution, and magnetic fields. The code we employ is arguably one of the most up-to-date and competitive codes in this regard, particularly in relation to treating fluid instabilities and mixing processes in astrophysical gases.


Our work stands apart from the papers referenced above in that it simulates a Galactic setup in three dimensions, rather than following an individual cloud or selection of cloudlets. The outflow is generated self-consistently from SF and subsequent SNe explosions within the Galactic disc, and we follow the Galactic fountain cycle over a period of Gyrs.

\section{Method}\label{sec:method}

Our simulations are run with our own modified version of the public \Gizmo code \citep{Hopkins2017c}. \Gizmo is a mesh-free hybrid between Lagrangian and Eulerian solvers, designed to benefit from the advantages inherent to each. It discretizes the equations of hydrodynamics on a set of discrete tracers which represent unstructured cells, and assigns the volume of the cells based on a kernel function. The flux between the cells is determined by solving the Riemann problem across the boundary, avoiding the need for artificial dissipation terms.

The hybrid mesh-free approach taken by \Gizmo allows for the mixing of different fluid phases naturally (as the flux of all fluid quantities is calculated between cells as default), while at the same time retaining the ability for Lagrangian adaptivity of cells and spatial resolution, and maintaining exact mass, energy, and momentum conservation. The meshless-finite-mass (MFM) approach that we use is also second-order consistent, which allows for formal convergence without needing to tend to the limit of an infinite neighbour number.


Radiative cooling in \Gizmo is performed down to $10$ K using the cooling tables of \cite{KWH1996}. The cooling prescription includes a uniform extragalactic ultra-violet (UV) background \citep{HaardtMadau2012}. We include full physical thermal conduction with a fiducial conduction coefficient of $0.1$ of the Spitzer value \citep{Spitzer1962}, and physical viscosity set to the Braginskii viscosity \citep{Hopkins2017b}. We also employ turbulent diffusion of metals across cells as per the Smagorinsky turbulent eddy diffusion model \citep{HopkinsEtal2017a, ColbrookEtal2017}. The gas has self-gravity in addition to an external analytic potential (see next Section).



\begin{figure*}
\centering
\includegraphics{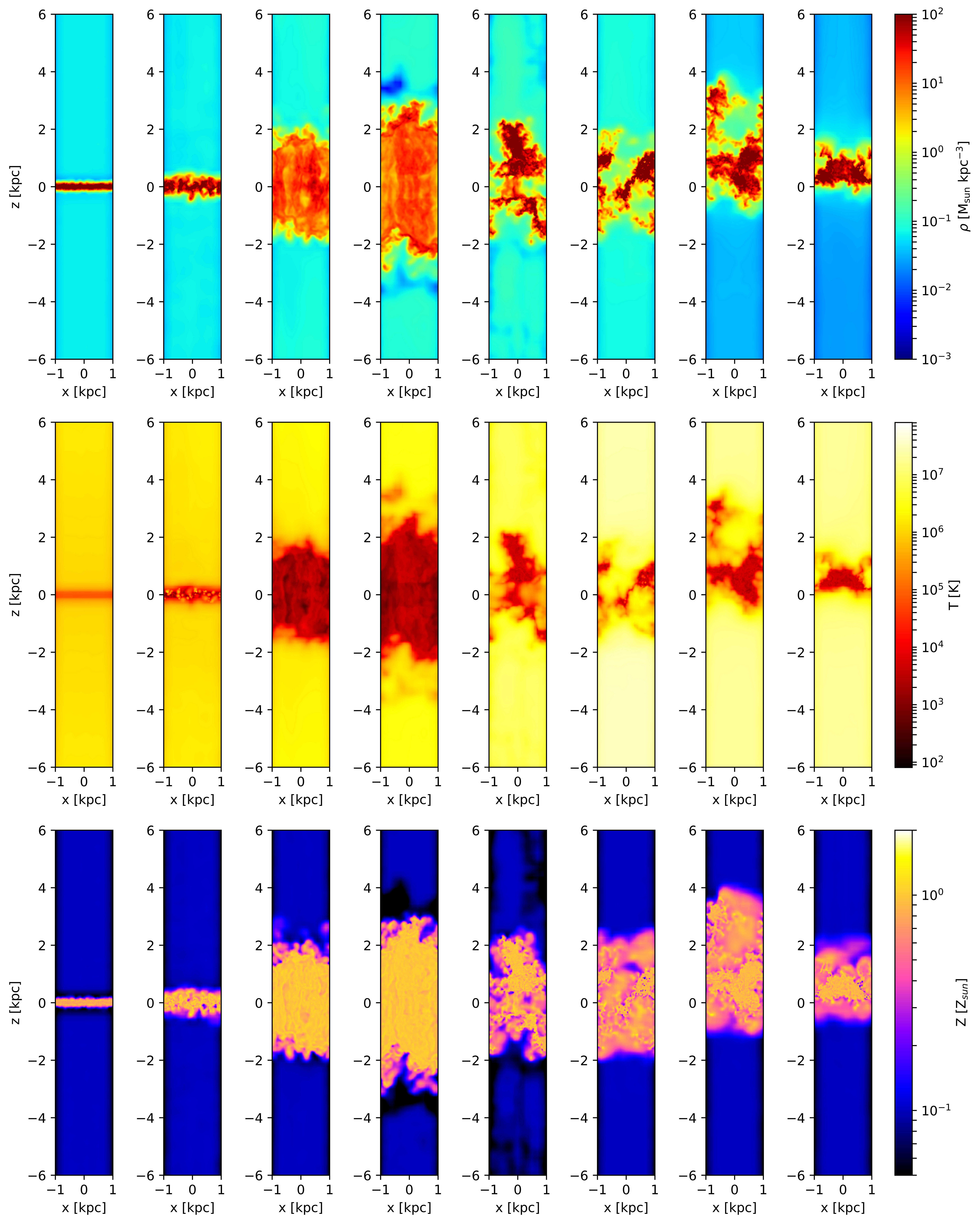}
\hspace{0.1in}
\caption[]{Visualisation for our fiducial SF10-turb run, showing density (top), temperature (middle), and metallicity (bottom) in the side-on view at 8 different times: $t = 1, 3, 10, 100, 200, 300, 400$, and $500$ Myr, left-to-right.}
\label{fig:vis90}
\end{figure*}

\subsection{Initial conditions}\label{sec:ic}

\begin{table*}
\centering
\caption{Summary of the simulations in this paper. $N_{\rm gas}$ is the number of gas particles, $m_{\rm gas}$ is the (constant) mass of each gas particle, $\eta_{\rm Sp}$ is the thermal conductivity in units of the Spitzer conductivity, $\nu_{\rm Brag}$ is the physical viscosity in units of the Braginskii viscosity, and $\rho_{\rm SF}$ is the threshold density for star formation in atoms per cc.}

\vspace{0.12in}
\begin{tabular}{|c c c c c c c|}\hline
\textbf{Simulation ID} & \textbf{$N_{\rm gas}$} & \textbf{$m_{\rm gas}$ ($\msun$)} & \textbf{$\eta$ ($\eta_{\rm Sp}$)} & \textbf{$\nu$ ($\nu_{\rm Brag}$)} & \textbf{$\rho_{\rm SF}$} (atoms per cc) & Turbulent IC \\ \hline

\hline
& & & & & &\\
H1 & $2.2 \times 10^6$ & $23$ & $0.1$ & $1.0$ & $10.0$ & Yes \\
H1-sf5 & $2.2 \times 10^6$ & $23$ & $0.1$ & $1.0$ & $5.0$ & Yes\\
H1-sf1 & $2.2 \times 10^6$ & $23$ & $0.1$ & $1.0$ & $1.0$ & Yes\\
H1-sf0.1 & $2.2 \times 10^6$ & $23$ & $0.1$ & $1.0$ & $0.1$ & Yes\\
H1-sf0.05 & $2.2 \times 10^6$ & $23$ & $0.1$ & $1.0$ & $0.05$ & Yes\\
H1-sf0.02 & $2.2 \times 10^6$ & $23$ & $0.1$ & $1.0$ & $0.02$ & Yes\\
H1-sf0.01 & $2.2 \times 10^6$ & $23$ & $0.1$ & $1.0$ & $0.01$ & Yes\\
& & & & \textbf{} & & \\

H1-$\eta1.0$ & $2.2 \times 10^6$ & $23$ & $1.0$ & $1.0$ & $10.0$ & Yes\\
H1-$\eta0.01$ & $2.2 \times 10^6$ & $23$ & $0.01$ & $1.0$ & $10.0$ & Yes \\
H1-$\nu0.1$ & $2.2 \times 10^6$ & $23$ & $0.1$ & $0.1$ & $10.0$ & Yes \\
H1-$\nu0.01$ & $2.2 \times 10^6$ & $23$ & $0.1$ & $0.01$ & $10.0$ & Yes \\
& & & & \textbf{} & & \\

M1 & $1.3 \times 10^6$ & $38$ & $0.1$ & $1.0$ & $10.0$ & Yes\\

L1 & $6.6 \times 10^5$ & $77$ & $0.1$ & $1.0$ & $10.0$ & Yes\\


& & & & & & \\

H2 & $2.2 \times 10^6$ & $23$ & $0.1$ & $1.0$ & $10.0$ & No \\

& & & & & & \\

\hline

\hline
\hline
\end{tabular}
\begin{flushleft}
\end{flushleft}
\label{tableics}
\end{table*}


Our model consists of a section of the MW disc, embedded in a hot ambient medium of low-density halo gas. The disc section has constant surface density in $x$ and $y$, and a Gaussian profile in the vertical direction $z$. We use periodic boundary conditions in the $xy$ plane, with an open boundary in the vertical ($z$) direction, and position the disc plane at the centre of the box, so that the halo gas is simulated both above and below the plane. This allows the model to cater for (i) differences in the outflow that is generated above and below the plane, and (ii) the effect on the outflow gas if any passes back through the disc and out the other side. A visualisation of the setup at an early time is shown in the left-most panel of Figure 1.

The setup was chosen based on the model in the SILCC papers \citep{SILCCpaper}, which approximates a present-day MW disc, although our implementation is $16$ times larger in the disc plane in order to explore a larger range of behaviour. The initial conditions (ICs) comprise a periodic box of gas measuring $2$ kpc $\times$ $2$ kpc $\times$ $\pm 50$ kpc (the extreme length in the $z$-direction is to ensure that any numerical effects at the open boundary affect the behaviour of the gas near the disc as little as possible). The disc has a midplane density of $9 \times 10^{-24}$ g cm$^{-3}$ and a scaleheight of $60$ pc. The disc vertical extent is cut off at the height where the disc density becomes equal to the ambient density, $\vert z \vert \approx 0.24$ kpc. The disc gas has solar metallicity. The total mass of the disc gas is $4 \times 10^7 \msun$. The halo gas has a density of $0.0006$ atoms cm$^{-3}$ and a metallicity of $0.1$ solar. In addition to the gas disc, a stellar disc analytic potential is included which has the form of a sech$^2$ profile -- the `isothermal sheet' \citep{Spitzer1942}, which is described by equation 29 in \cite{SILCCpaper}. The stellar disc potential has a surface density of $30 \msun$ pc$^{-2}$ and a scaleheight of $100$ pc.

Our IC is set up to be initially in hydrostatic equilibrium with both the gas distribution and the external analytic gravitational potential. The temperature in the disc varies in the vertical direction, $T \equiv T(z)$, from $\sim 10^1 - 10^5$ K (the majority being at $\approx 5 \times 10^2$ K). The temperature in the halo (within $10$ kpc) is $T \approx 10^6$ K. 

For most of the ICs (including the fiducial) we seed a one-off, initial turbulent velocity field in the disc that is constructed based on the treatment in \cite{DubinskiNarayanPhillips1995}. Briefly, the velocity field follows a Kolmogorov power spectrum,
\begin{equation}
P_{v}(k) \sim k^{-11/3},    
\end{equation}
where k is the wavenumber. Such a velocity field is homogeneous and incompressible, and so we can define $\bf{v}$ in terms of a vector potential $\bf{A}$ such that $\bf{v} = \bf{\nabla} \times \bf{A}$. The components of $\bf{A}$ can then be described by a Gaussian random field with an associated power spectrum,
\begin{equation}
P_{A}(k) \sim k^{-17/3}.    
\end{equation}
The variance in $\vert \bf{A} \vert$ diverges sharply as $k$ decreases, and so we introduce a small-scale cut-off $k_{\rm min}$. We can then write:
\begin{equation}
<\vert {\bf{A_k}} \vert^2 > = C(k^2+k_{\rm min})^{-17/6},
\end{equation}
where C is a constant that sets the normalization of the velocities. We set this equal to unity and normalize the velocity field once the statistical realization has been generated. Physically, the small-scale cut-off $k_{\rm min}$ can be interpreted as the scale $R_{\rm max} \simeq k_{\rm min}^{-1}$ , the largest scale on which the turbulence is likely to be driven. In our model we use $R_{\rm max} = 1$ kpc.

To generate the statistical realization of the velocity field we sample the vector potential $\bf{A}$ in Fourier space, drawing the amplitudes of the components of $\bf{A_k}$ at each point ($k_x$, $k_y$, $k_z$) from a Rayleigh distribution with a variance given by $⟨\vert \bf{A_k} \vert^2 ⟩$ and assigning phase angles that are uniformly distributed between $0$ and $2\pi$. We then take the curl of $\bf{A_k}$,
\begin{equation}
{\bf{v_k}} = i \bf{k} \times \bf{A_k},
\end{equation}
to obtain the Fourier components of the velocity field. Finally we take the Fourier transform of this to obtain the velocity field in real space. We use a periodic cubic grid of dimension $256^3$ when generating the statistical realization of the velocity field and we use tri-cubic interpolation to estimate the components of the velocity field at the position of each particle/cell.

Although the setup of our model is based on the SILCC model \citep{SILCCpaper}, our focus and science goals differ greatly. The SILCC papers deal with the state of the Galactic interstellar medium (ISM) for the purposes of understanding SF within the disc; we are interested in the mixing of disc gas and halo gas far from the disc in order to understand positive feedback. Our implementation also stands apart by employing a meshless hydro solver. 


With the small section of the MW disc that we model, we are able to achieve a considerably higher resolution than if we modelled the entire Galaxy -- in our fiducial runs, this is $23 \msun$ per cell. The resolution of all the runs is shown in Table 1.

\subsection{Star formation and feedback}\label{sec:sf}

Star formation in our simulations proceeds subject to a density threshold that differs by 3 orders of magnitude ($0.01-10$ atoms cm$^{-3}$) between different runs and the molecular hydrogen fraction criterion of \cite{HopkinsEtal2017a, KrumholzGnedin2011}. The latter criterion is particularly useful at SF density thresholds $< 100$ atoms cm$^{-3}$ in correcting the lack of self-shielding molecules at low resolution. Feedback from stars is modelled according to \cite{AGORApaper}, and injects thermal energy and metals from an IMF-averaged number of type II SNe in a single burst after a main sequence time. The injected energy is weighted over the kernel. We note that this is a relatively simple SF and feedback model, but it has been shown to be effective in high resolution simulations with particle masses $< 100 \msun$ \citep{HopkinsEtal2018}. As listed in Table 1, the majority of our simulations have a mass resolution of $23 \msun$, are therefore able to make effective use of this model.








\section{Results}\label{sec:results}

\begin{figure*}
\centering
\includegraphics[width=1.0\textwidth]{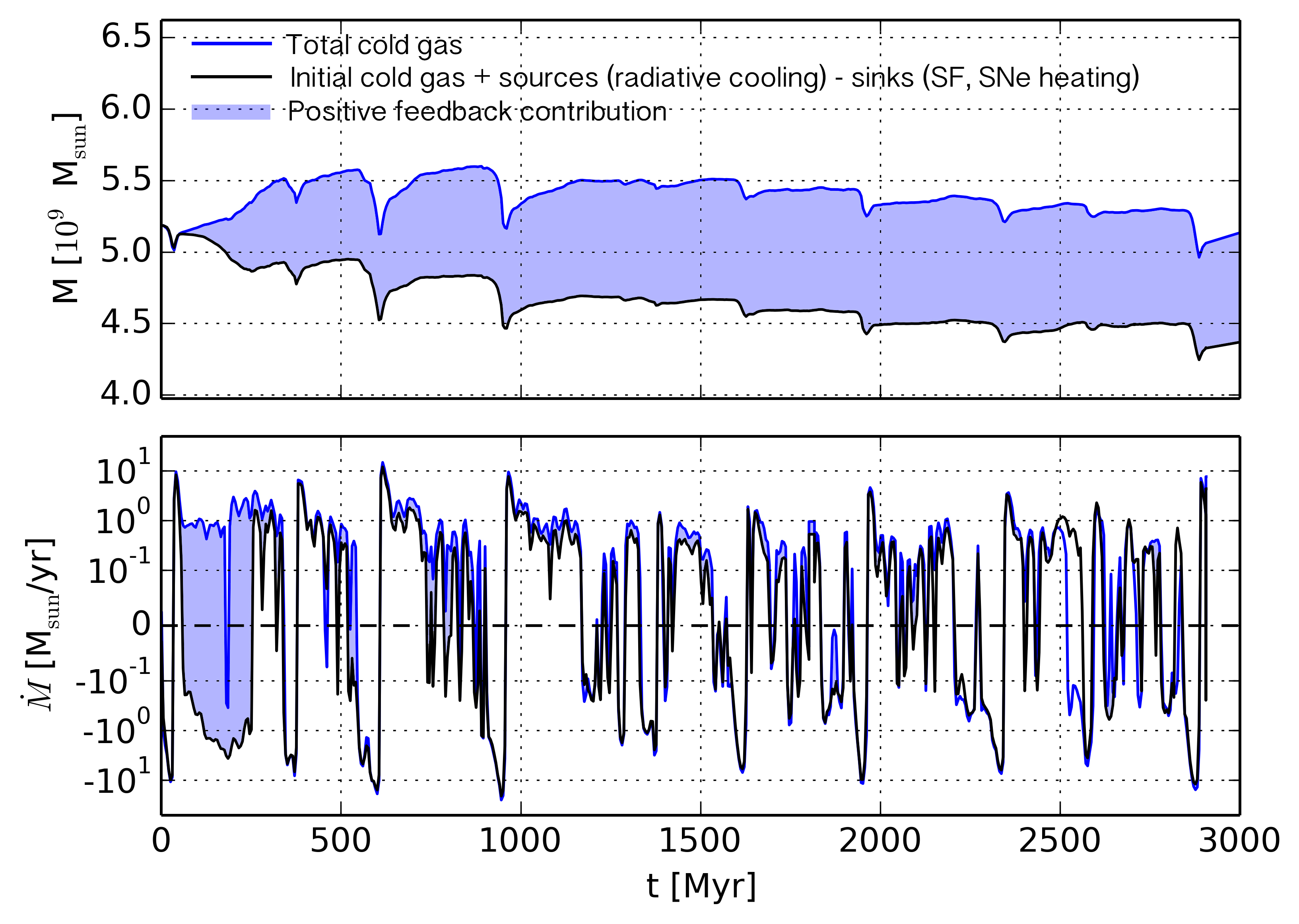}
\hspace{0.1in}
\begin{minipage}[b]{1.0\textwidth}
\caption[]{Total amount of cold gas mass in the entire simulation over $3$ Gyr (top) and the corresponding time derivative (bottom). The top panel compares the total cold gas (blue line) with the amount of cold gas that there would be assuming no contribution from the `halo gas extraction' positive feedback process -- namely, the black line is the original amount of cold gas that is gradually used up by SF and heated by SNe, which also includes the (very small) increase from radiative cooling of the initial hot ambient gas. The grey shaded area therefore represents the contribution to the cold gas budget made by positive feedback.}
\end{minipage}
\label{fig:coldtotal}
\end{figure*}

In Figure 1 we show a time-series with the state of the fiducial (H1) model at 8 different times: $t = 1$ Myr, $t = 3$ Myr, $t = 10$ Myr, $t = 100$ Myr, $t = 200$ Myr, $t = 300$ Myr, $t = 400$ Myr, and $t = 500$ Myr. In this Figure we show the density (top), temperature (middle) and metallicity (bottom), all averaged along the line-of-sight.

The initial star formation and subsequent SNe explosions taking place in the disc give rise to a lifting of some of the disc gas out to $\sim 2-3$ kpc. The material is not ejected, however, since the outflow velocity is not high enough to overcome both the gravitational potential and the confining pressure of the hot gas. This results in the previously-lifted gas falling back and re-forming the disc, at which point it undergoes a subsequent starburst and the cycle repeats. This cycle repeats approx. every 300 Myr. The maximum height reached of the disc gas (the outflow) varies across the different runs, and is in particular sensitive to the strength of the initial star formation.

Each time the disk reforms, its structure and morphology are different, which naturally causes differences in the morphology of the SF and subsequent feedback. These differences compound as the simulation progresses, and so we note that these simulations are firmly in a somewhat chaotic regime whereby small changes between setups can amplify and give rise to large differences at later times. As a result we do not investigate morphologies in any detail, or make comparisons between runs at specific times; rather, we look at global, averaged quantities where possible.

The initial condition was varied between initially turbulent and initially not-turbulent, with very little difference found in the behaviour of the gas over the run (see Section \ref{sec:convic}). The one-off seeded turbulence, while playing a role in creating some structure and density variation in the disc, is ultimately less effective than the SNe themselves in determining the dynamics of the outflow.

\subsection{Cold gas: mass increase and distribution}

\textbf{Note:} For all quantitative plots and recorded values we extrapolate from our $2$ kpc $\times$ $2$ kpc section of the MW disc to the entire disc -- therefore, numbers quoted for the amount of cold gas, SF, the corresponding time derivatives, and all relevant figures, have been multiplied by $M_{\rm disc}/M_{\rm patch}$, where $M_{\rm disc}$ is the total mass of the present-day MW gaseous disc, and $M_{\rm patch}$ is the total mass of the section that we model. The conversion factor is $M_{\rm disc}/M_{\rm patch} = 132.5$. The division by $M_{\rm patch}$ gives us each relevant quantity per unit mass (in other words, a `specific' rate for each quantity), which we then scale up to the whole MW disc.
This simplistic scaling is not meant as a substitute for a full scale modelling of a MW disk but rather as a means to facilitate a comparison of our results with observations. For instance, this scaling will clearly overpredict the time variation of the global SFR given that in reality different patches may experience a burst at different times.


\vspace{0.1in}

\noindent The nature of positive feedback from SF is that the action of SNe actually gives rise to an \emph{increase} in cold gas, rather than a decrease. As such, we track the amount of cold gas in the simulation over time, where `cold gas' is defined as having $T < 2 \times 10^4$ K \citep[as per][]{ArmillottaEtal2016}. This is shown in Figure 2 for our H2 run over 3 Gyrs (blue line). The amount of cold gas in the simulation is largely constant over the course of the 3 Gyrs, even with vigorous SF converting cold gas to stars and SNe heating the surrounding gas to high T. The SF in the simulation over the 3 Gyrs is shown in Figure 3. From the bottom panel of this Figure, which shows the SFR, one can see that the SF is bursty, alternating between periods of high SFR (reaching values of $\simgt 1 \msun$ yr$^{-1}$) before shutting off and waiting for the disc to re-form. The near-constant cold gas mass is therefore indicative of a `self-sustaining' MW + hot halo ecosystem, whereby cold gas sent into the hot halo gas is able to mix and entrain the hot gas and cool it, adding to the reservoir of fuel for SF the next time the cycle repeats.

Of course, in such a model it can be difficult to separate out the relevant processes from each other and determine the contribution that each makes. To show the contribution from positive feedback alone, we have plotted in Figure 2, alongside the total cold gas mass, the amount of cold gas that would have been seen in the simulation should this positive feedback process not have added to the cold gas budget. This is the black line in Figure 2, and it was calculated by taking the initial cold gas mass and (i) subtracting the SF and SNe heating over the lifetime of the simulation (the cold gas `sinks') and (ii) adding the radiative cooling of the hot halo gas (a cold gas `source'). The difference between the black and the blue lines on this plot (the shaded blue area) is therefore the contribution from positive feedback alone. It is clear that the vertical height of this shaded area grows quickly in the first 400 Myr, after which it remains largely the same. We discuss this initial vs. latter phase and the reasons for it in Section \ref{sec:effect}.


\begin{figure*}
\centering
\includegraphics[width=1.0\textwidth]{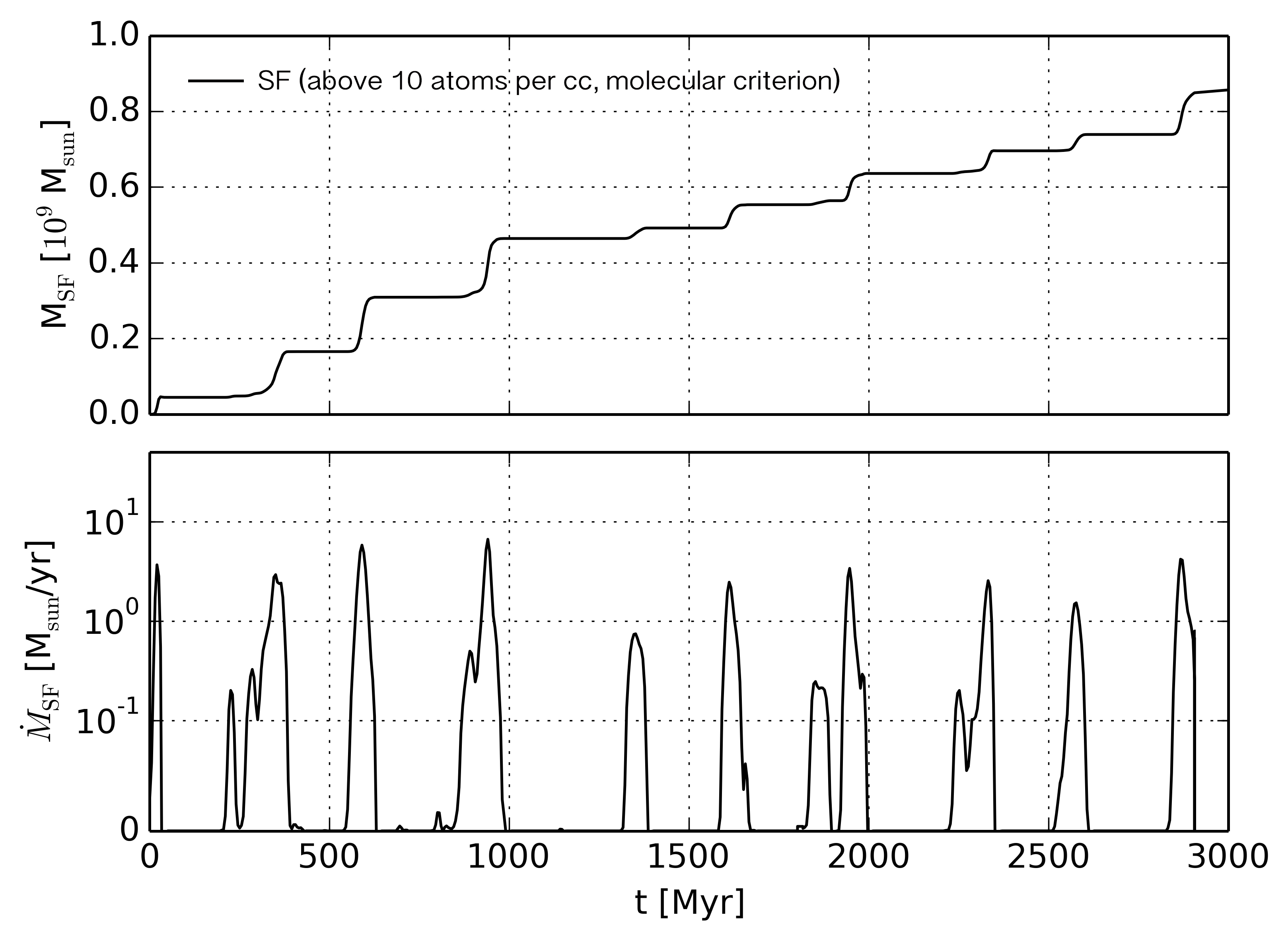}
\hspace{0.1in}
\begin{minipage}[b]{1.0\textwidth}
\caption[]{Total amount of mass in newly-formed stars in the entire simulation over $3$ Gyr (top) and the corresponding rate (bottom). Here is it clear that we have a Galactic fountain cycle repeating many times, with each peak in the SFR (bottom panel) corresponding to the disc re-forming before a subsequent starburst that blows the gas out for $\approx 200-300$ Myrs, during which time there is very little to no SF.}
\end{minipage}
\label{fig:sftotal}
\end{figure*}

\subsection{Positive vs. negative feedback}


To further investigate this process of positive feedback from SF, and in particular, to compare it to the usual negative feedback that is caused by the heating of the gas through SNe, we tracked the hot halo gas and the disc gas separately through their IDs. We define a metric for positive feedback as gas that was originally in the hot halo phase achieving a temperature below $2 \times 10^4$ K \citep[temperature threshold defined as per][]{ArmillottaEtal2016}. We also define a metric for negative feedback; gas that was originally in the disc reaching a temperature above $2 \times 10^5$ K (this was originally the highest temperature of the gas that made up our Gaussian disc IC). In this way, we can measure how much of the initially-hot gas became cold, and how much of the initially-cold gas became hot. Since the cooling time for gas $> 10^6$ K is long, we can be confident that most of the cooled hot halo gas is the result of the disc gas mixing with the halo gas (and indeed, we have verified this by running a stand-alone model with only the hot ambient gas allowed to cool radiatively with no influences from the disc). The initially-cold gas in the disc that gets heated is the result of the SNe injecting thermal energy into their surroundings.

\begin{figure*}
\centering
\includegraphics[width=1.0\textwidth]{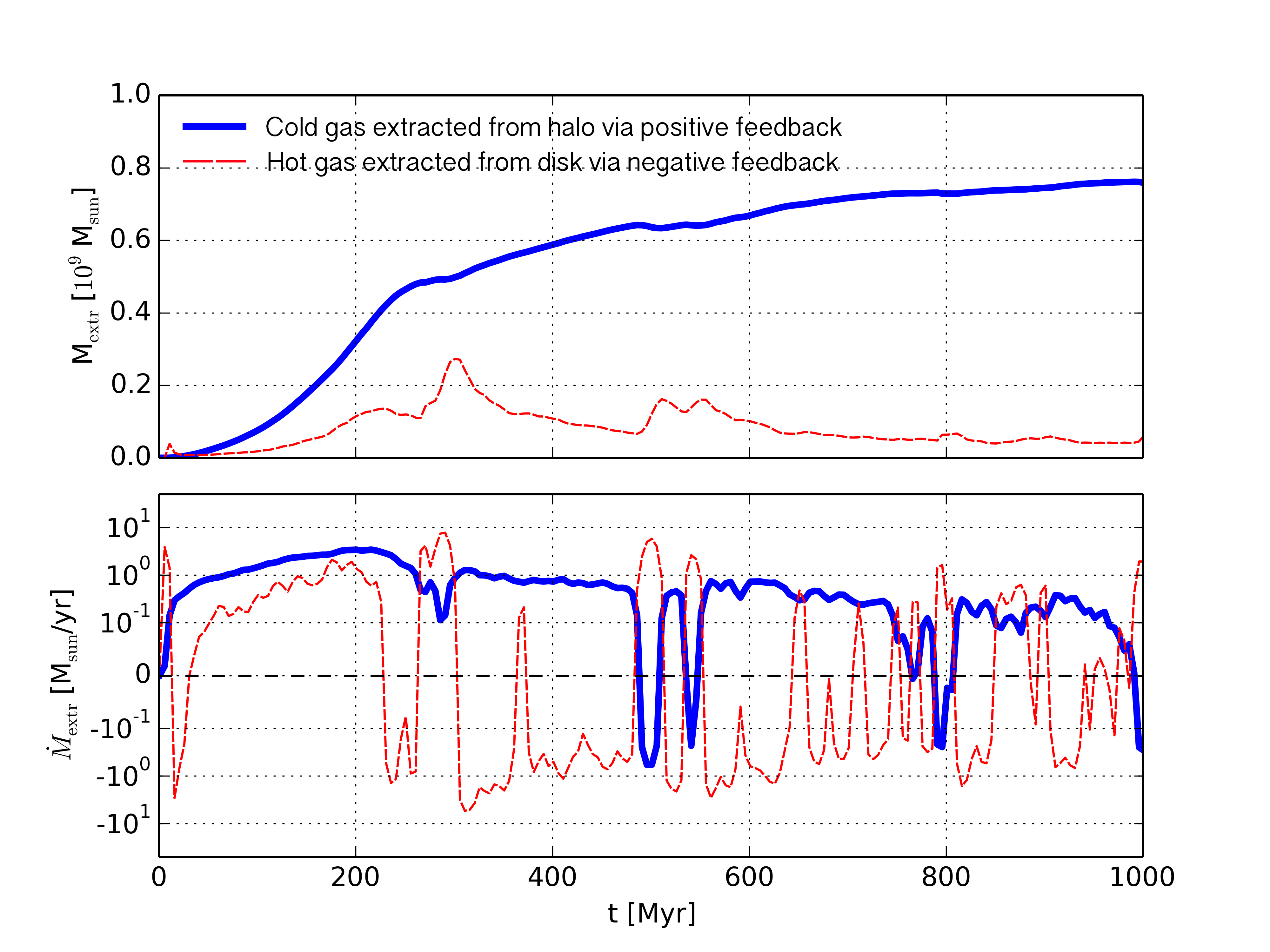}
\hspace{0.1in}
\begin{minipage}[b]{1.0\textwidth}
\caption[]{Positive vs. negative feedback comparison for H1 (fiducial). Extracted gas mass (top panel) and the corresponding time derivative (bottom panel). The cold gas extracted from the hot halo is several times larger than the hot gas extracted from the disc for the majority of the run. The reason for this can be seen in the bottom panel, where although both the cold and hot extracted gas reach similar peaks and troughs, the extracted cold gas rate stays mostly in the upper (positive) half while the extracted hot gas rate oscillates almost equally above and below zero.}
\end{minipage}
\label{fig:hotcold}
\end{figure*}

Figure 4 shows these two metrics over $1$ Gyr for our H1 (fiducial) run. The top panel shows the total mass of `extracted' gas, and the bottom panel shows the time derivative. The initially-disc gas that is heated up is naturally somewhat bursty, since the SNe heating coincides with the semi-regular starbursts \citep[see, e.g.,][]{StinsonEtal2007}. The cold gas extracted from the hot halo (via positive feedback) grows faster than hot gas created from the disc, showing that positive feedback metric exceeds negative feedback in this setup. The rate-of-change of the cold/hot gas extracted from the halo/disc differs quite clearly, with the extracted cold gas rate remaining mostly above zero (with occasional dips) and the extracted hot gas oscillating above and below zero far more often. The peak values of $\dot{M}_{\rm extr}$, however, are similar.




\subsection{Varying the SF threshold}

Since both the positive feedback and the negative feedback mechanism depend on SF, it makes sense to explore varying SF rates. While the SF rate is a quantity that arises naturally through the behaviour of the cold gas in the simulation, and cannot be set directly by hand, we can use the SF threshold (listed in Table 1 for each run) as a proxy. In varying the density threshold at which stars are allowed to form (although still subject to the molecular gas fraction criterion) we can achieve different average SF rates across the lifetime of the simulation. The different SF thresholds give rise to different amounts of cold gas extracted from the halo via the positive feedback process. These are shown in Figure 5. The different runs produce roughly the same amount of extracted cold gas (top panel) and the corresponding time derivative (bottom panel) up to $\approx 200$ Myr, after which they begin to diverge. This point in time coincides roughly with the second starburst and the first repetition of the fountain cycle, and indeed small differences in the initial amount of star formation can lead to far larger differences each time the cycle is repeated. In general, the rates of cold gas extraction for each of the varying-SF threshold runs have a similar form post 200 Myr, but the locations and frequency of the dips below zero vary, with the lowest SF threshold runs having the most dips into negative $\dot{M}_{\rm extr}$.

Figure 6 shows the mass of stars formed (top panel) and the SFR (bottom panel) for this suite of runs varying the SF threshold. A lower SF threshold naturally leads to a larger starburst at $t = 0$, although the peak value reached saturates at $0.1$ atoms per cc. The major difference however is clearly seen in how bursty the SFR is across the different runs. 
At a high SF threshold ($10$ atoms per cc run in our fiducial run), SF is restricted to high density gas and, thus, can only take place in the gas disk. In contrast, at a low SF threshold, stars are able to form more easily in outflowing gas and at higher altitudes above the midplane. This has the effect of creating a more continuous SF history over the run, although the typical maximum SFR values reached are similar. However, the more sustained SFR for the thresholds at the lower end of range result in a total stellar mass that is $\approx 4$ times higher than our fiducial 10 atoms per cc run. 
In general, a higher SF threshold returns a lower average SF rate across 1 Gyr, as seen in table 2. The relationship is not proportional, however, as the behaviour of the gas in the simulation is extremely chaotic and influenced by many different factors. 

From Figure 5 we see that, above $\rho_{\rm SF} = 1$ atom cm$^{-3}$, the amount of extracted cold gas decreases as the SF threshold is increased. Below $\rho_{\rm SF} = 1$ atom cm$^{-3}$, this trend is reversed, with a higher SF threshold leading to \emph{more} cold gas extracted from the halo. This is an interesting result, and, given the inverse scaling between SFR and SF threshold, it leads us to offer an alternative definition of positive feedback; that of a positive feedback \textbf{trend}, whereby an increase in SFR gives rise to an increase in the cold gas extraction rate. Similarly, we define a `negative feedback trend' as one in which an increase in SFR in the Galaxy leads to a decrease in the rate of cold gas extraction (as is the usual interpretation of the effect of SNe on the Galactic cold gas budget). The bottom panel of Figure 5 shows the rate of cold gas mass extracted from the hot halo, and from this we can get an average rate over the run. These values are listed in Table 2 for each value of the SF threshold. In Figure 6 we see the mass of SF (top panel) and the corresponding rate (bottom panel) for the same set of runs. We can extract an average SF rate for each run from this Figure, and these are also listed in Table 2.

Nonetheless, there is a general trend of lower average SFR with higher SF threshold. We therefore have 7 different values of the average SFR over 1 Gyr, which we can plot against the average rate of cold gas extraction over the same period. This is shown in Figure 7. When we do this, we find that at SFRs below $\approx$ 0.5 $\msun$ yr$^{-1}$, the extracted cold gas rate and thus the cold gas budget for the Galaxy increases with increasing SFR. This demonstrates in a simple plot our 'positive feedback trend' below $0.5 \msun$ yr$^{-1}$. Above this SFR value, we see that the behaviour of the extracted cold gas rate with SFR follows a negative feedback trend -- increasing the SFR above $0.5 \msun$ yr$^{-1}$ extracts less and less cold gas from the hot halo. This Figure will be discussed in detail in Section \ref{sec:discussion}, as it has interesting implications for galaxy evolution if we are to assume that positive feedback processes play a significant role.

The next plot, Figure 8, shows the same two positive feedback/negative feedback regimes but with the average rates drawn from either $t < 400$ Myr or $t > 400$ Myr, since it is clear from all the previous plots vs. time that the behaviour of the outflow in terms of extracting cold gas is significantly different within the first $400$ Myr compared to later times. As we remark on in Section \ref{sec:discussion}, this is not due to an initial transient arising from the IC, but rather a consequence of the favourable conditions for cold gas extraction within the first $400$ Myr of the simulations. In Figure 8 we see indeed that the positive feedback trend peaks at a higher value, $\approx 2 \msun$ yr$^{-1}$, for $t < 400$ Myr than when averaged over the whole Gyr. 
The trend for $t > 400$ Myr also shows a peak near ${\rm SFR}\sim{}0.5\,M_{\odot}\,{\rm yr}^{-1}$, although the peak is much less pronounced and the cold gas extraction rate shows only a weak scaling with SFR for low SFRs. 
This explains why we see only weak effects of positive feedback after $t = 400$ Myr. 

\begin{figure*}
\centering
\includegraphics[width=1.0\textwidth]{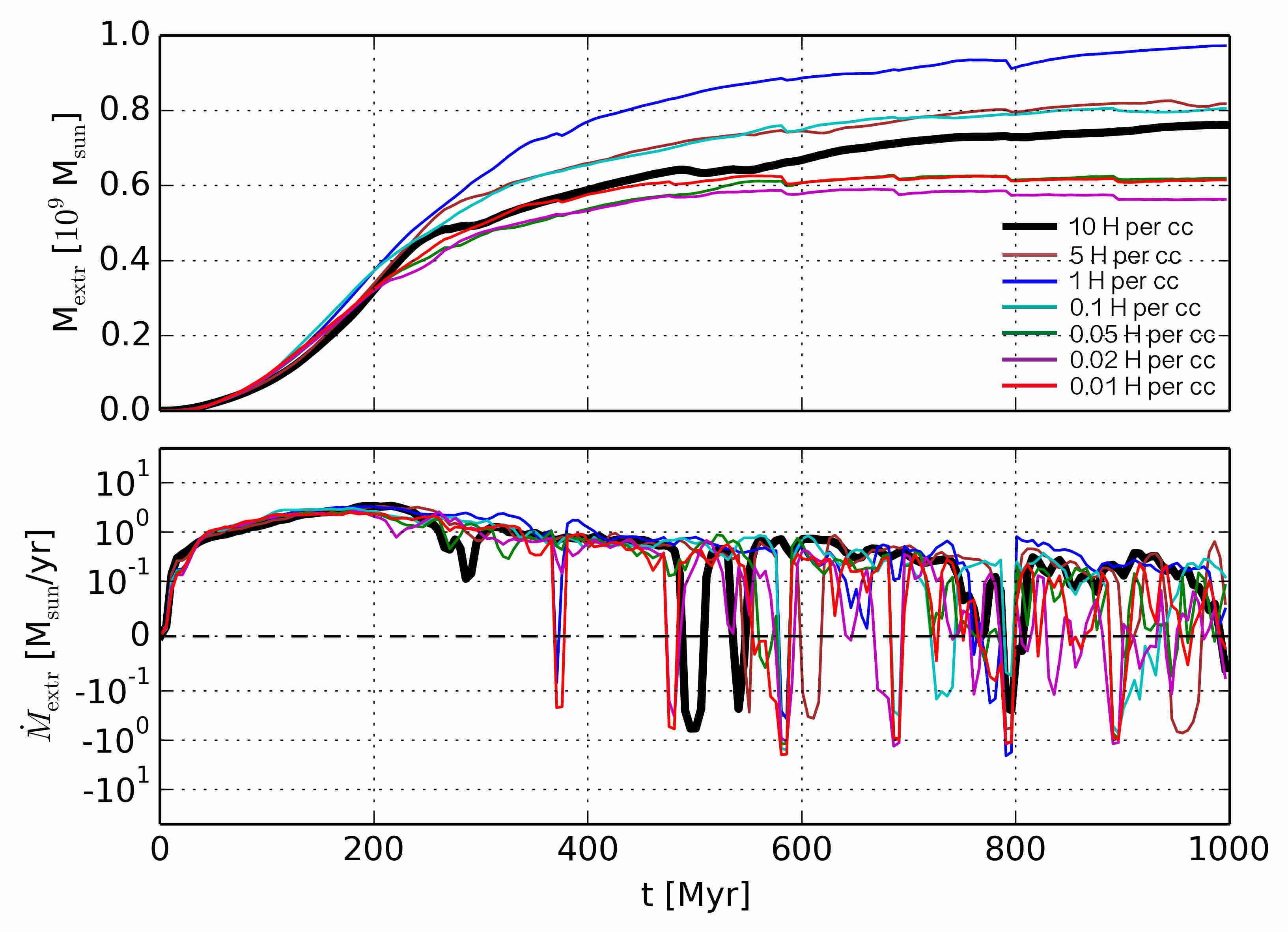}
\hspace{0.1in}
\begin{minipage}[b]{1.0\textwidth}
\caption[]{Plot showing the mass of extracted cold gas (top) and the corresponding time derivative (bottom), comparing simulations H1-sf with SF thresholds as listed in the key. The fiducial H1 run is shown with a thick black line for clarity. For the high SF thresholds, a reduction in the threshold leads to a greater amount of cold gas by the end of the run; however this trend reverses as the SF threshold drops below $1$ atom cm$^{-3}$. For the very low SF thresholds (0.01, 0.02 atoms cm$^{-3}$, the amount of extracted cold gas starts to slowly decrease again towards the end of the run.}
\end{minipage}
\label{fig:coldmulti}
\end{figure*}

\begin{figure*}
\centering
\includegraphics[width=1.0\textwidth]{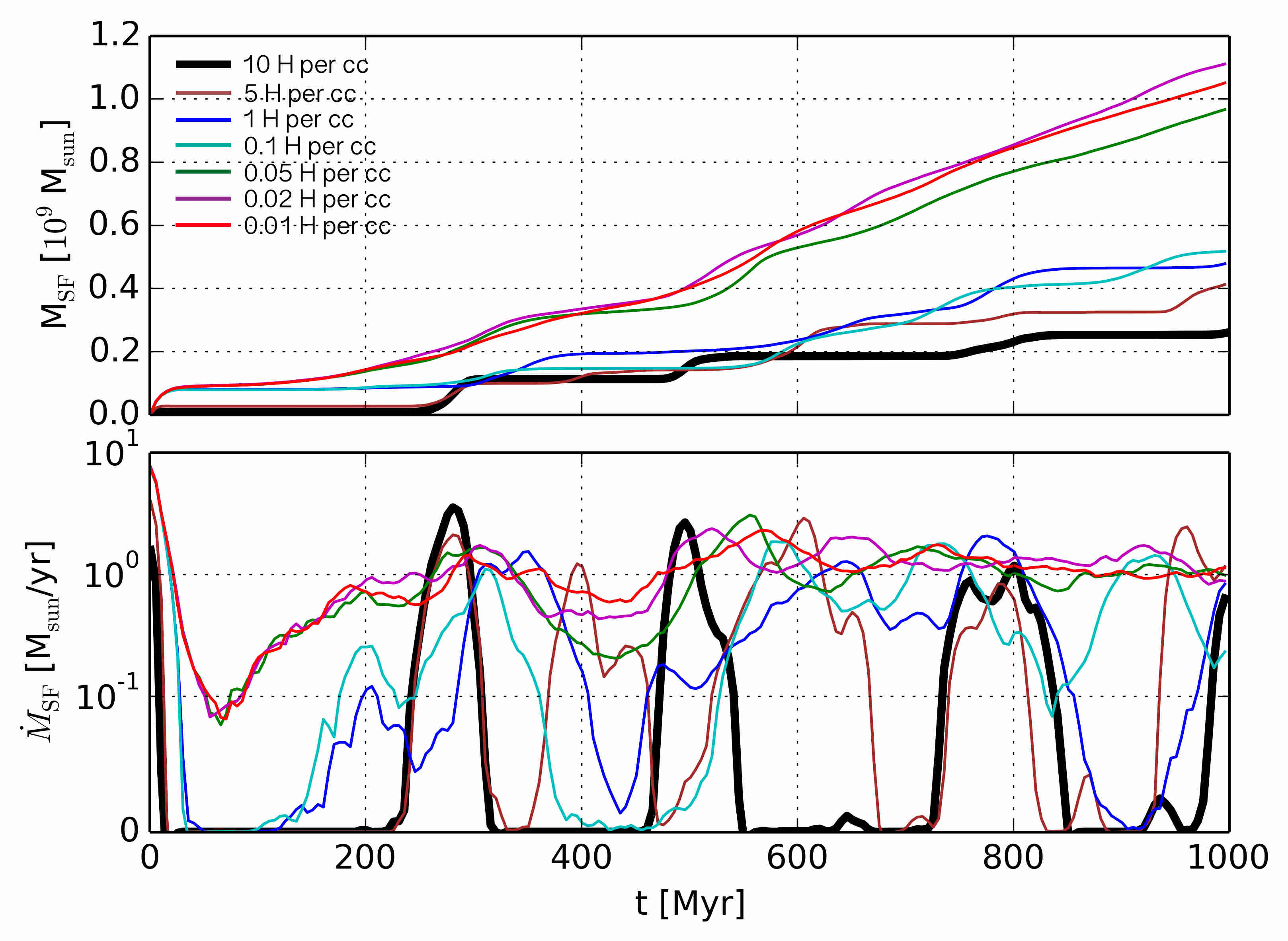}
\hspace{0.1in}
\begin{minipage}[b]{1.0\textwidth}
\caption[]{Plot showing the mass of stars formed (top) and the SFR (bottom), comparing simulations H1-sf with SF thresholds as listed in the key. The fiducial H1 run is shown with a thick black line for clarity. In general, the high SF thresholds give bursty SF histories, while the low SF thresholds show a more continuous, high SFR over the course of the run.}
\end{minipage}
\label{fig:sfmulti}
\end{figure*}

\begin{table}
\centering
\caption{$\dot{M}_{\rm cold}$ is the average rate of `cold' gas mass increase (defined as $T < 2 \times 10^4 K$) and $\dot{M}_{\rm hot}$ is the average rate of hot gas mass increase (defined as $T > 2 \times 10^5 K$) in the simulation over $1$ Gyr. $\dot{M}_{\rm SF}$ is the average SF rate over $1$ Gyr. The final two columns provide average mass conversion rates; $\dot{M}_{\rm cold,x}$ is the average rate of cold gas ($T < 2 \times 10^4 K$) extracted from the initial hot ambient gas, and $\dot{M}_{\rm hot,x}$ is the average rate of hot gas ($T > 2 \times 10^5 K$) formed from the disc gas. All values in this table are in units of $\msun$ yr$^{-1}$.}
\vspace{0.12in}
\begin{tabular}{|c c c c c c c|}\hline
\textbf{ID} & \textbf{$\dot{M}_{\rm cold}$} & \textbf{$\dot{M}_{\rm hot}$} & \textbf{$\dot{M}_{\rm SF}$} & \textbf{$\dot{M}_{\rm cold,x}$} & \textbf{$\dot{M}_{\rm hot,x}$} \\ \hline

\hline
& & & & \textbf{} & \\
H1 & $0.462$ & $-0.873$ & $0.262$ & \textbf{0.761} & $0.051$ \\
H2 & $0.147$ & $-0.791$ & $0.465$ & \textbf{0.717} & $0.130$ \\
& & & & \textbf{} &  \\

H1-sf5 & $0.311$ & $-0.886$ & $0.415$ & \textbf{0.811} & $0.105$ \\

  &      &     &     &     &  \\
H1-sf1 & $0.464$ & $-1.095$ & $0.480$ & \textbf{0.966} & $0.059$ \\

 &      &     &     &     &  \\
H1-sf0.1 & $0.273$ & $-0.951$ & $0.517$ & \textbf{0.804} & $0.052$ \\
&      &     &     &     &  \\
H1-sf0.05 & $-0.432$ & $-0.801$ & $0.969$ & \textbf{0.611} & $0.123$ \\

 &      &     &     &     &  \\
H1-sf0.02 & $-0.627$ & $-0.743$ & $1.113$ & \textbf{0.555} & $0.129$ \\
 &      &     &     &     &  \\
 
H1-sf0.01 & $-0.517$ & $-0.769$ & $1.053$ & \textbf{0.604} & $0.127$ \\
 &      &     &     &     &  \\
 
\hline

\hline
\hline
\end{tabular}
\begin{flushleft}
\end{flushleft}
\label{tablesf}
\end{table}

\begin{figure*}
\centering
\includegraphics[width=1.0\textwidth]{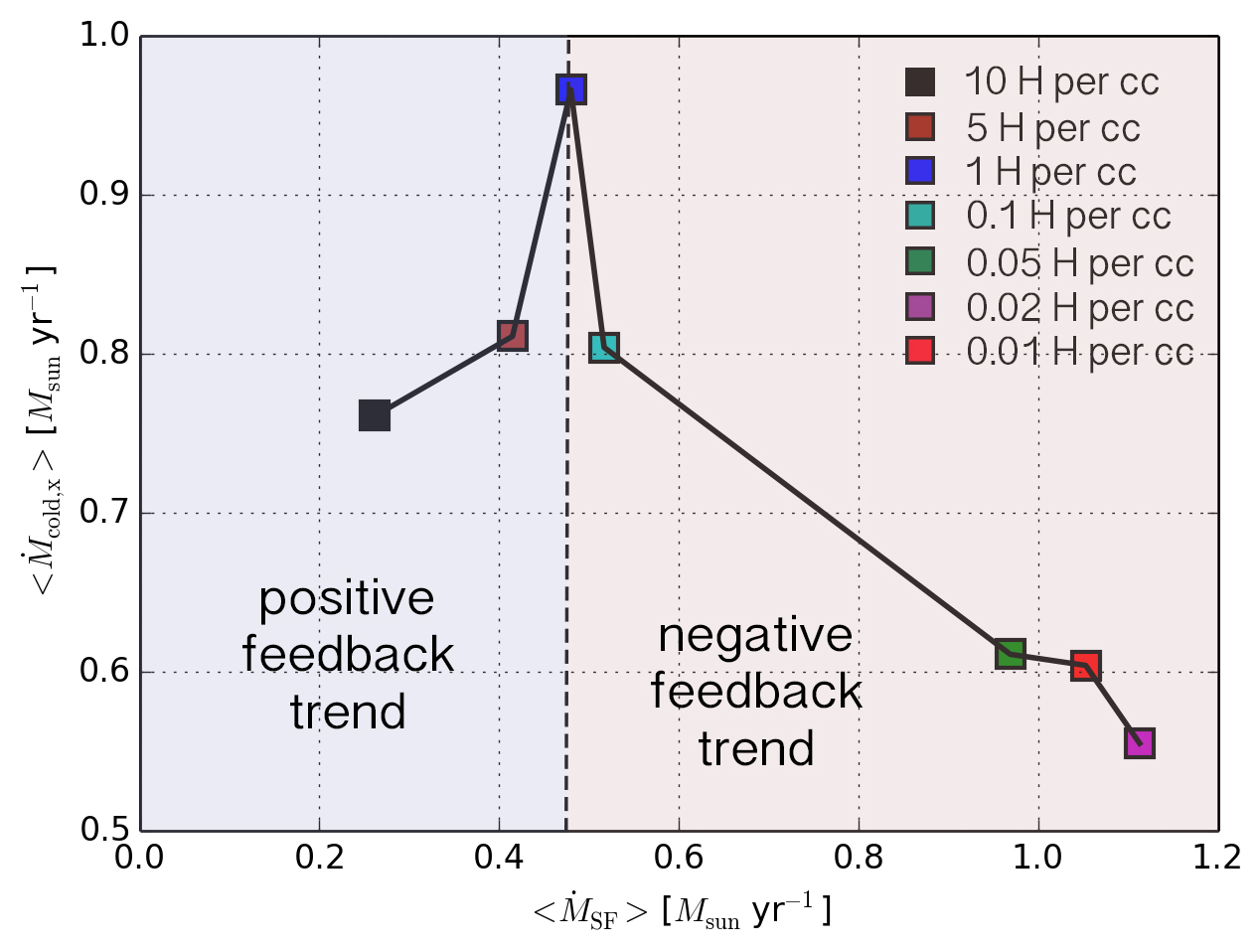}
\hspace{0.1in}
\begin{minipage}[b]{1.0\textwidth}
\caption[]{Trend of extraction rate of cold gas mass with SFR, both averaged over 1 Gyr, for the H1-sf runs with SF thresholds as shown in the key. The low average SFRs show a trend of increasing average cold gas mass rates with increasing SFR, while the high SFRs show the opposite. We have therefore made a distinction between a `positive feedback trend' and a `negative feedback trend', with the peak being approx. $0.5 \msun$ yr$^{-1}$.}
\end{minipage}
\label{fig:sftrend}
\end{figure*}

\begin{figure*}
\centering
\includegraphics[width=1.0\textwidth]{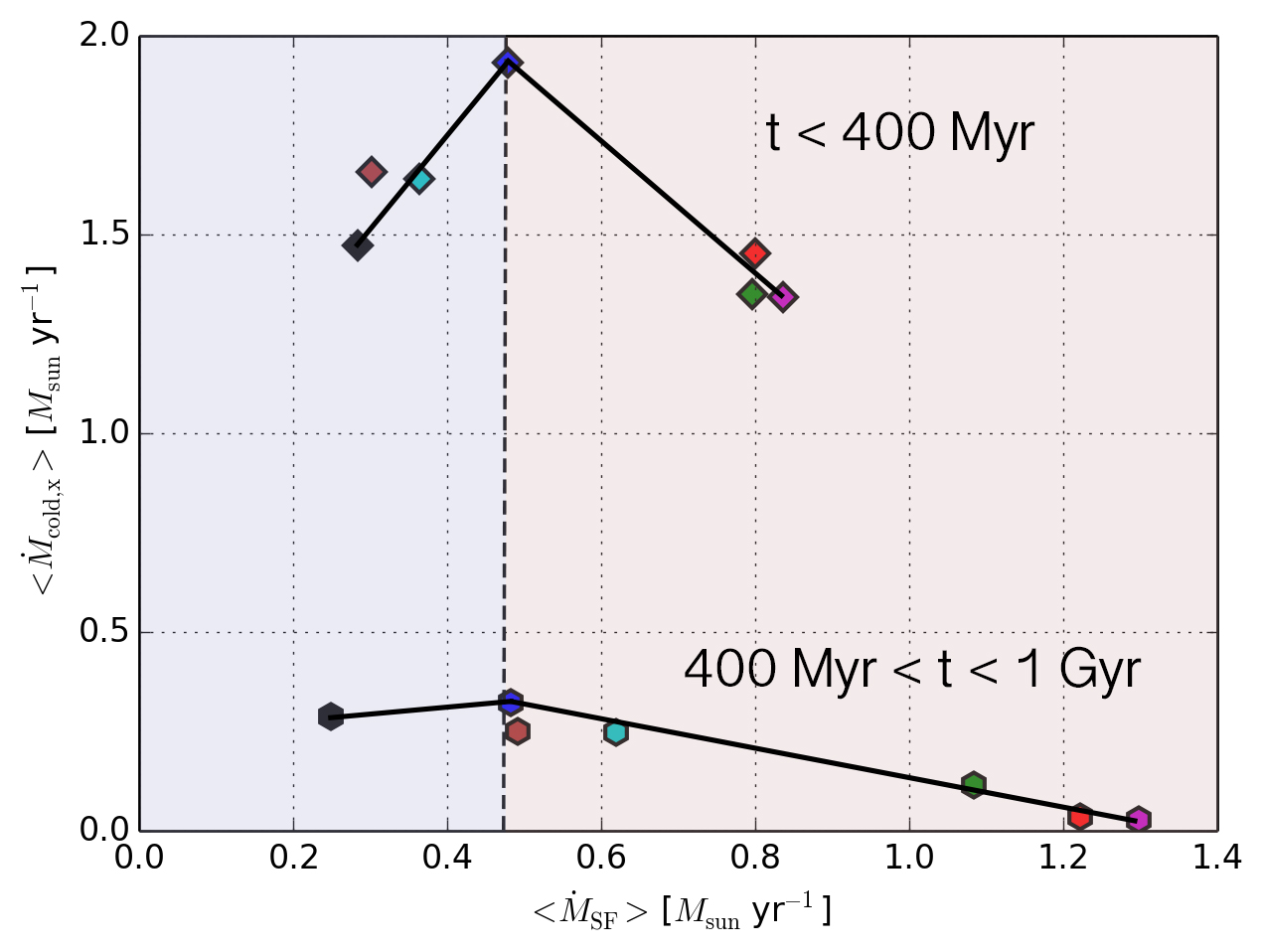}
\hspace{0.1in}
\begin{minipage}[b]{1.0\textwidth}
\caption[]{Average rate of extracted cold gas mass trend with average SFR, for the H1-sf runs with SF thresholds as shown in the key in Figure 7. Here we take the average in two parts; for $t < 400$ Myr (diamonds), and for $400$ Myr $< t < 1000$ Myr (pentagons). Note: lines connecting the symbols are purely instructive; to indicate an approximate trend (or not).}
\end{minipage}
\label{fig:sftrend2}
\end{figure*}

\subsection{Varying the model: turbulence, physical parameters, and resolution}\label{sec:conv}


In this section we compare the results for the extracted cold gas mass and SF (and the corresponding time derivatives) over 1 Gyr while varying numerical aspects of the model to test for convergence.

\subsubsection{Varying the IC}\label{sec:convic}

\begin{figure}
\centering
\includegraphics[width=0.5\textwidth]{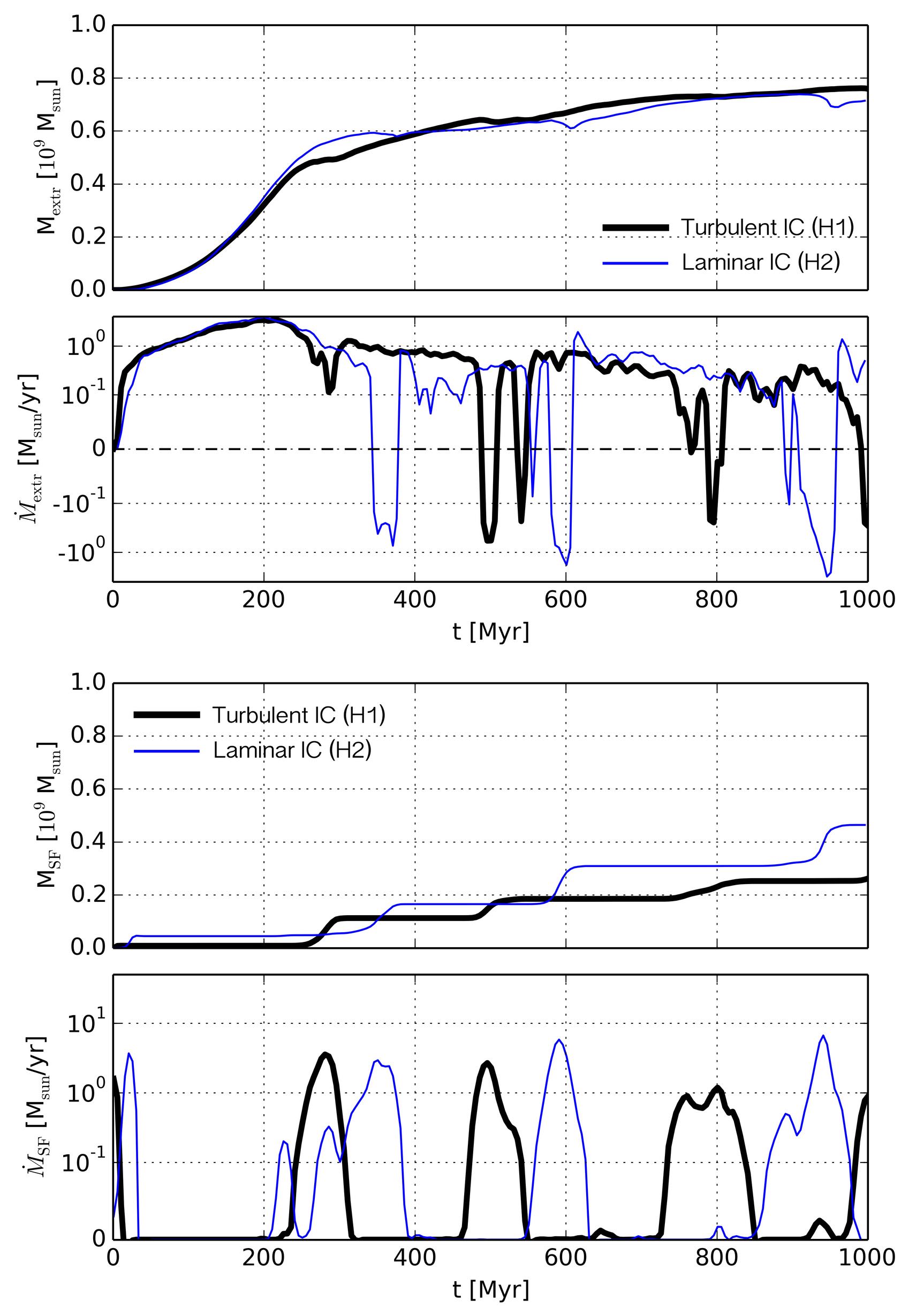}
\hspace{0.1in}
\begin{minipage}[t]{0.5\textwidth}
\caption[]{Comparison between extracted cold gas (top) and SF (bottom) for H1 and H2 - the fiducial run with the turbulent IC and the run with the non-turbulent IC.}
\end{minipage}
\label{fig:turbnoturb}
\end{figure}

The first is the setup of the initial condition -- we have run two different ICs, one with a turbulent velocity field in the disc (see Section \ref{sec:ic}) and one without. This is important for determining to what extent the details of the IC play a role in how the outflow gas extracts and cools gas from the halo. We find that there is very little difference between the two setups when it comes to the mass of cold gas extracted from the ambient hot gas or the stellar mass as shown in Figure 9. The rates of each differ only in terms of the location of the peaks/troughs, due to the somewhat chaotic nature of the simulation and the susceptibility of the starburst cycle to small changes in the parameters.  

\subsubsection{Varying the resolution}

\begin{figure}
\centering
\includegraphics[width=0.5\textwidth]{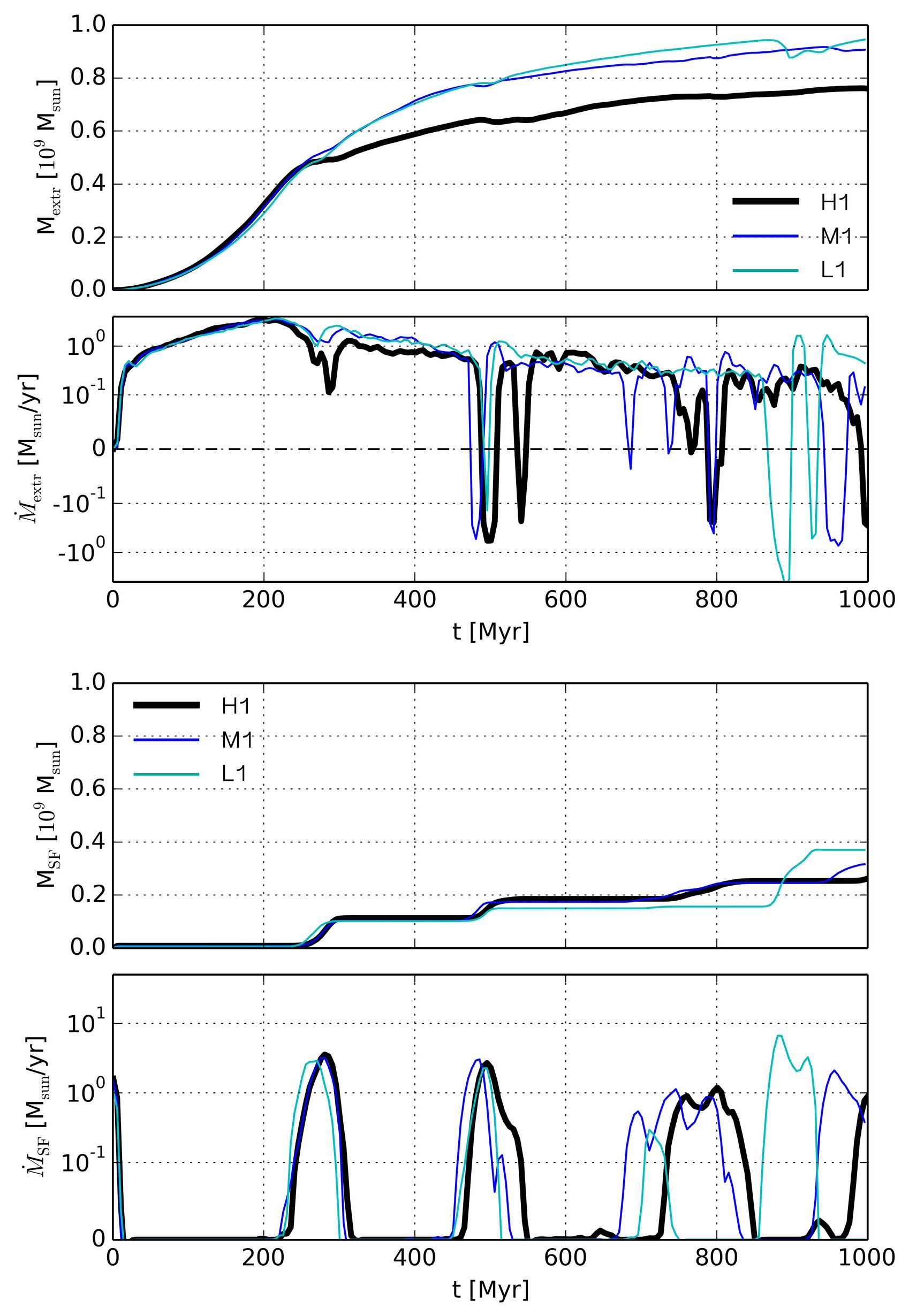}
\hspace{0.1in}
\begin{minipage}[t]{0.5\textwidth}
\caption[]{Comparison between three different resolutions -- H1, M1 and L1 (see Table 1) for the extracted cold gas from the hot ambient medium (top) and the star formation history (bottom) over 1 Gyr.}
\end{minipage}
\label{fig:resmulti}
\end{figure}

Figure 10 shows the extracted cold gas and SF over 1 Gyr for 3 different resolutions, as listed in Table 1 (H1, M1, and L1 runs). These were all performed with our fiducial, turbulent IC. The two lower resolution simulations produce slightly more cold gas by the end of the run at 1 Gyr, but this is mostly due to a higher rate of cold gas increase for M1 and L1 between $\approx 300-500$ Myr, after which the gradients are roughly the same across all three resolutions.

\subsubsection{Varying the thermal conductivity}

\begin{figure}
\centering
\includegraphics[width=0.5\textwidth]{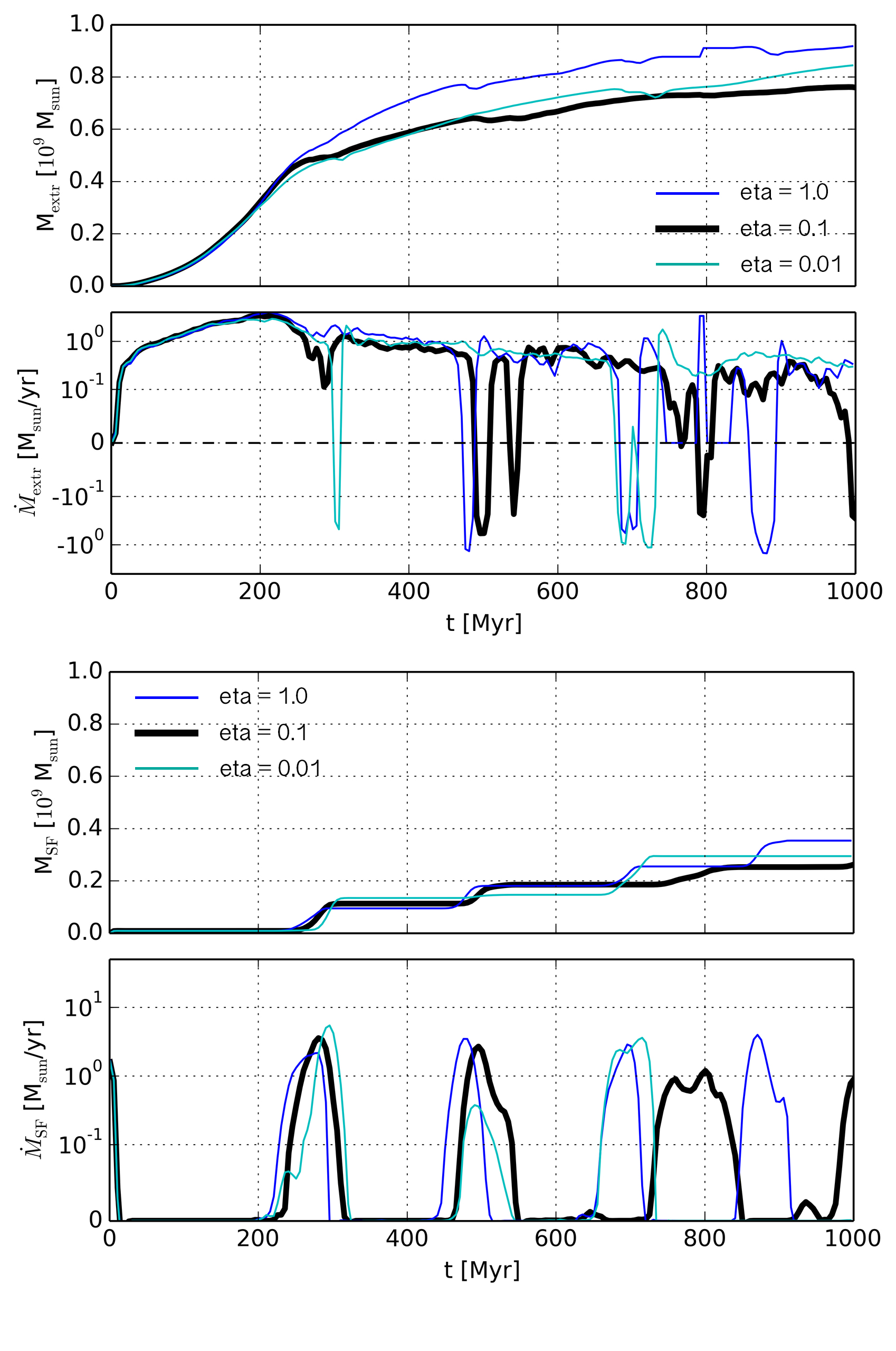}
\hspace{0.1in}
\begin{minipage}[b]{0.5\textwidth}
\caption[]{Comparison between three values of the thermal conductivity parameter, $\eta = 1.0$, $0.1$ (fiducial), $0.01$  $\eta_{\rm Sp}$ for the extracted cold gas from the hot ambient medium (top) and the star formation history (bottom) over 1 Gyr.}
\end{minipage}
\label{fig:condmulti}
\end{figure}


We also explore the effect of varying the thermal conductivity parameter, $\eta$, which refers to the fraction of the Spitzer conductivity that particles in the simulation possess. Figure 11 shows three values of $\eta$, 1.0, 0.1 (fiducial) and 0.01. There appears to be no obvious trend in how the thermal conductivity affects the amount of cold gas collected, or the mass of stars formed, and the total mass of extracted cold gas after 1 Gyr varies by $\approx 2 \times 10^8 \msun$, although the $\eta = 0.1$ and $\eta = 0.01$ runs are converged up to $400$ Myr. 

\subsubsection{Varying the physical viscosity}

\begin{figure}
\centering
\includegraphics[width=0.5\textwidth]{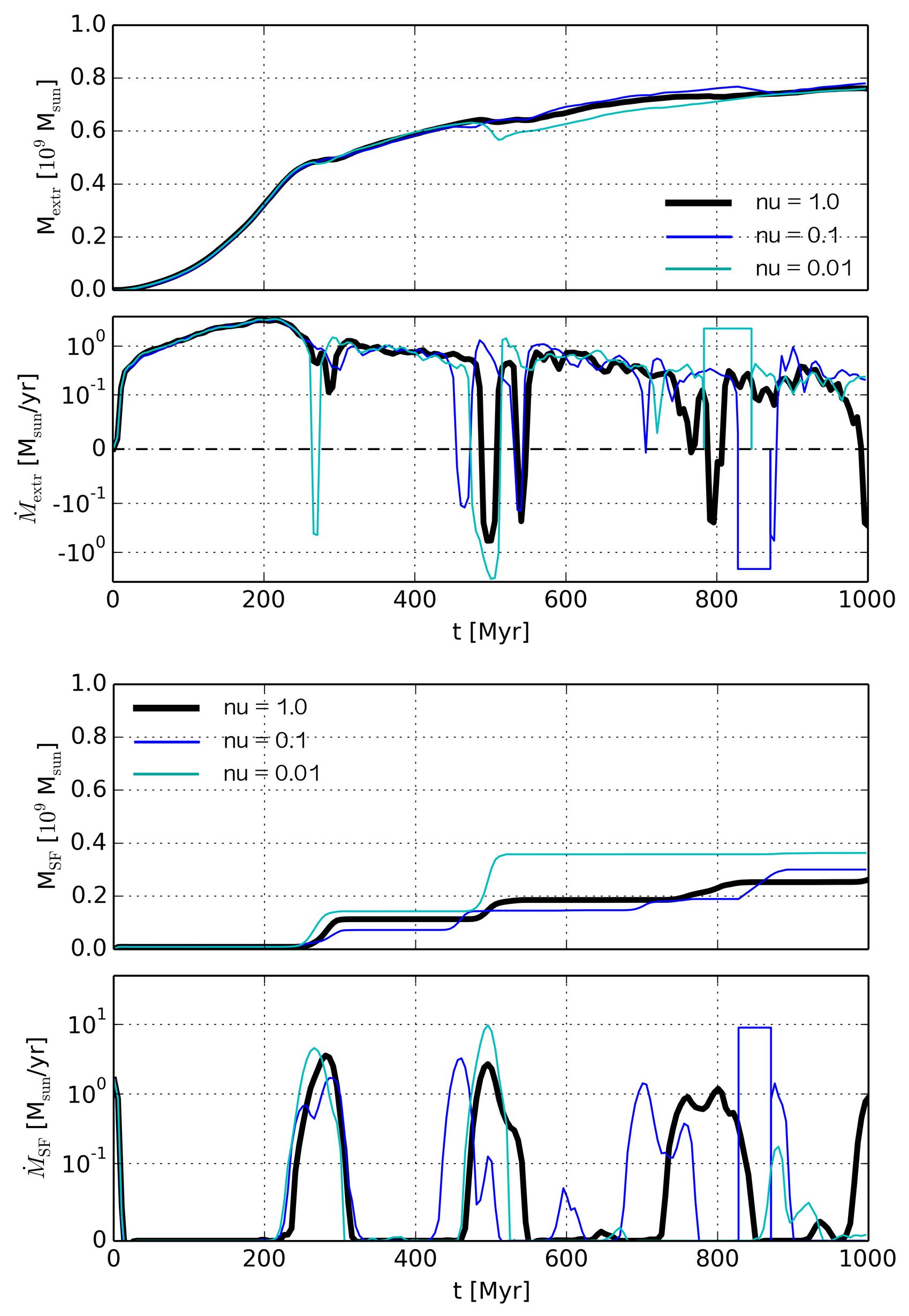}
\hspace{0.1in}
\begin{minipage}[b]{0.5\textwidth}
\caption[]{Comparison between three different values of the thermal conductivity parameter, $\nu = 1.0$ (fiducial), $0.1, 0.01$ $\nu_{\rm Brag}$ for the extracted cold gas from the hot ambient medium (top) and the star formation history (bottom) over 1 Gyr. Note: horizontally-flat sections in the rate correspond to missing data.}
\end{minipage}
\label{fig:viscmulti}
\end{figure}

Finally, we vary the normalisation of the physical viscosity $\nu$, as a function of the Braginskii viscosity. Figure 12 shows three values of $\nu$, 1.0 (fiducial), 0.1, and 0.01. The physical viscosity parameter seems to have very little effect on the total amount of cold gas collected by 1 Gyr, with the top panel very well converged. The corresponding time derivatives differ slightly in where the dips are but on average are similar. The star formation is converged but naturally differs after $t = 400$ Myr as to when the starbursts occur.

The purpose of this section was to compare the effect of varying a few of the parameters, but it should be noted that we kept the SF threshold fixed at the fiducial value of 10 atoms per cc. As such we are not investigating the SFR-cold gas extraction trend that we discuss in Section 4.2 and display in Figures 7 \& 8. Without running an exhaustive suite of simulations the effect of varying these parameters remains unknown -- it is possible that both the position and the normalisation of the peak separating the positive and negative feedback trends would be affected.


\section{Discussion}\label{sec:discussion}

\begin{figure*}
\centering
\includegraphics[width=1.0\textwidth]{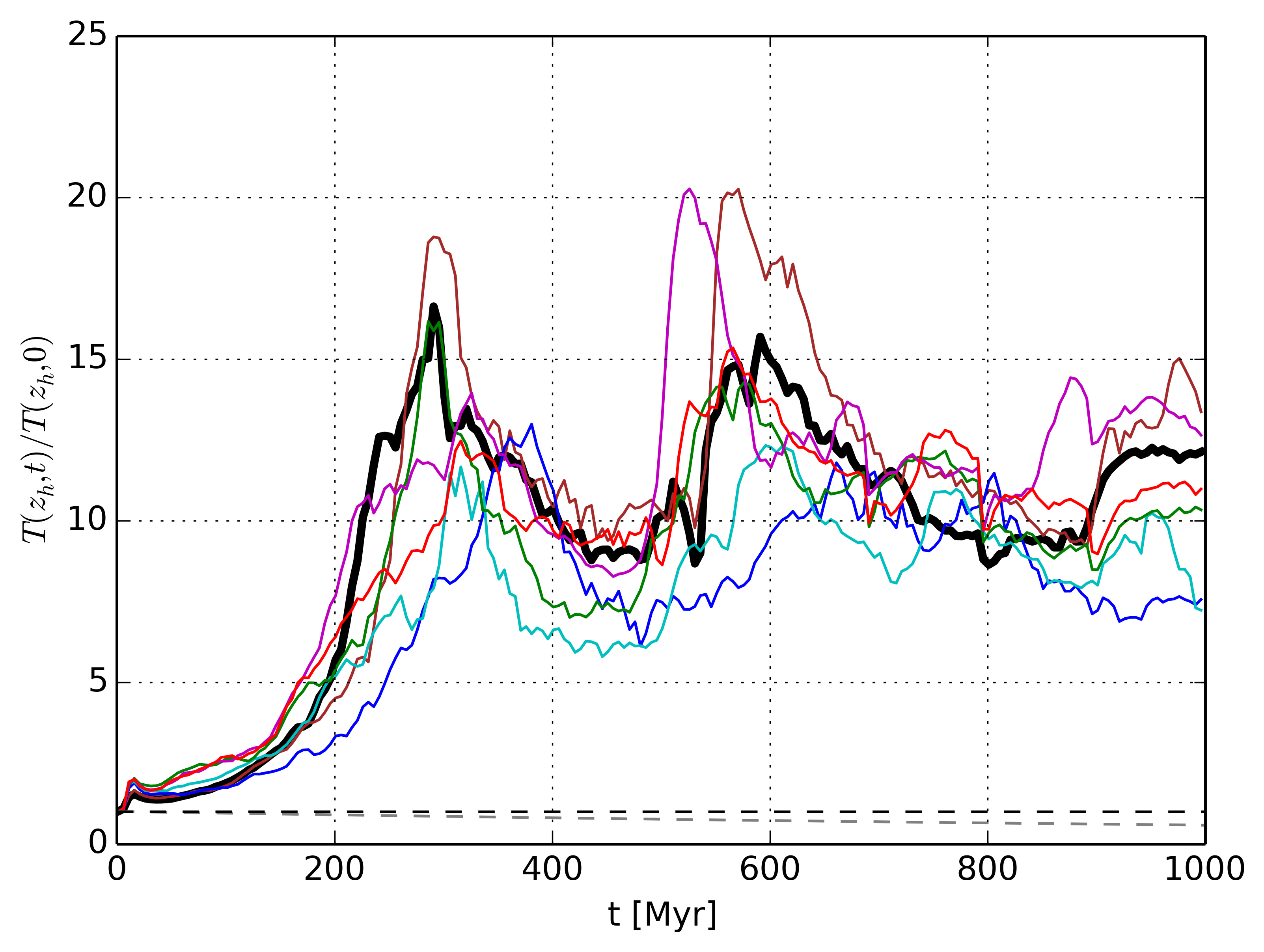}
\hspace{0.1in}
\begin{minipage}[b]{1.0\textwidth}
\caption[]{Plot showing the ratio of average temperature to initial temperature, for the hot ambient gas in a region between 0.5 and 1x the maximum height that the cold dense gas has reached by time $t$. Once again we compare simulations H1-sf with SF thresholds as listed in Figure 6. The fiducial H1 run is shown with a thick black line for clarity. We also include the average temperature of this region assuming ONLY the ambient halo gas (therefore any cooling is due strictly to radiative cooling losses), this is shown with a grey dashed line. The black dashed line marks the initial ratio of 1.}
\end{minipage}
\label{fig:Tmulti}
\end{figure*}

\begin{figure*}
\centering
\includegraphics[width=1.0\textwidth]{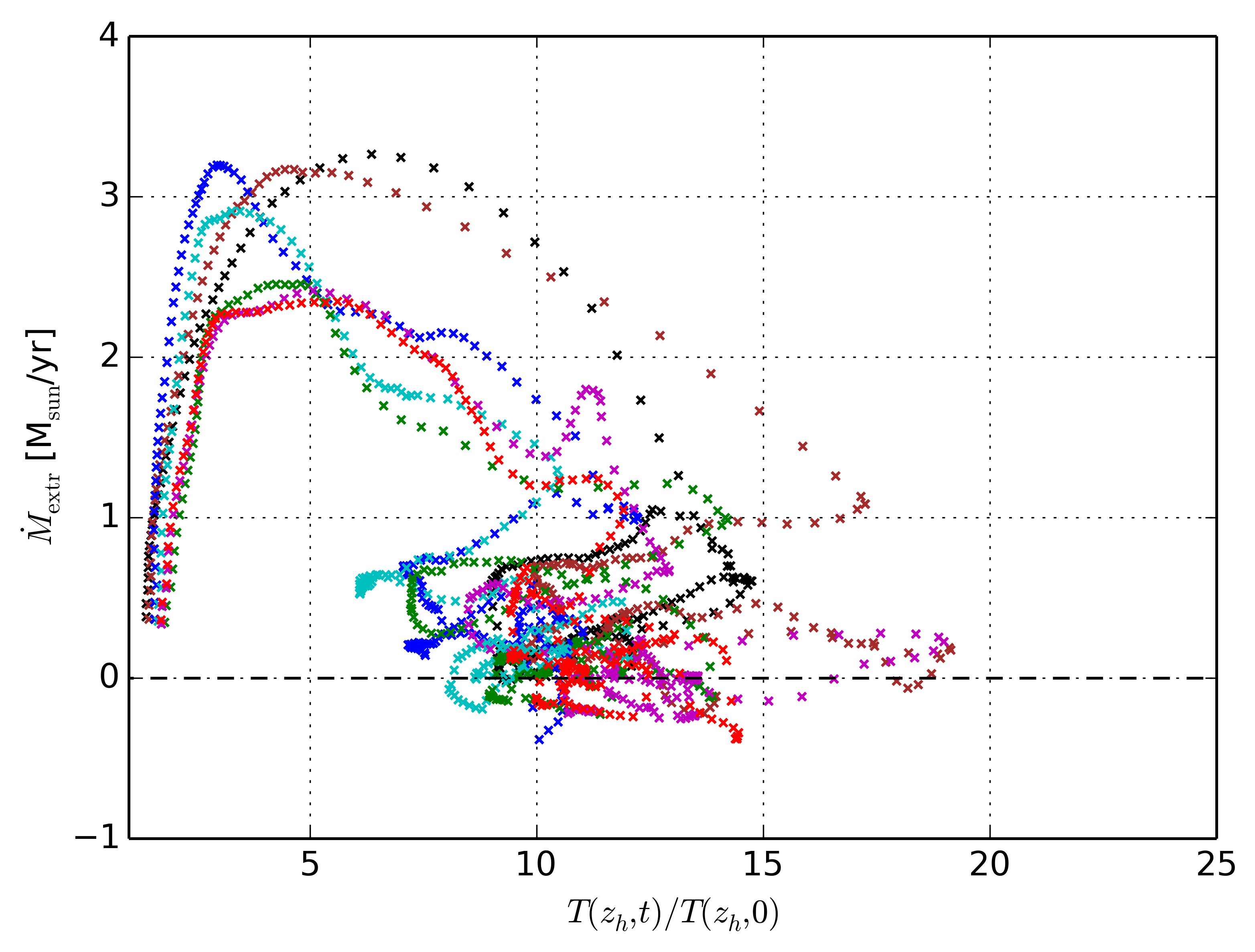}
\hspace{0.1in}
\begin{minipage}[b]{1.0\textwidth}
\caption[]{Plot showing the temperature ratio from Figure 13 (the ratio of average temperature to initial temperature) vs. the extracted cold gas rate from Figure 5, to give an idea of how the efficiency of cold gas extraction varies with increasing ambient temperature. For this plot we take the running mean values on each axis over a time interval of 50 Myr. Note: crosses that are closer together in x-axis spacing indicate that the run spends more time at those temperatures.}
\end{minipage}
\label{fig:Tvsmulti}
\end{figure*}

The action of SNe feedback promoting additional cooling is a departure from the conventional wisdom on the role of SNe outflows in galaxy formation and evolution. Typically, galaxies have been modelled as some form of `bathtub', with source and sink terms \citep{LillyEtal1996, DekelMandelker2014} where SNe outflows exclusively remove cold gas from a galaxy. This constitutes `negative feedback', which has been the accepted norm. Positive feedback, which is a somewhat more recent idea \citep{MarinacciEtal2011, FraternaliTomassetti2012, ArmillottaEtal2016, HobbsEtal2016}  complicates this picture by introducing, via a galactic fountain process operating in the hot gaseous halo, a means to increase the galaxy's store of cold gas available for star formation.

The mechanism by which this occurs is a mixing of cold gas initially located within the galactic disc with hot gas in the halo. The cold gas is sent into the hot halo gas by the action of SNe feedback, lifting it from the plane of the galaxy but not ejecting it from the galaxy's potential well, so that it later falls back after mixing with hot halo gas. This parcel of mixed hot and cold gas shows a net cooling, as the cooling rate at $\sim 10^6$ K (the approx. value for the hot halo gas that interacts with the fountain gas) is approx. an order of magnitude lower than the cooling rate at $10^5$ K. A parcel of gas loses energy $E$ at a rate,
\begin{equation}
\frac{dE}{dt} = -n^2 \Lambda(T),
\end{equation}
where $n$ is the number density and $\Lambda(T)$ is the cooling rate. The cooling time can be defined as
\begin{equation}
t_{\rm cool} \equiv \frac{E}{\vert dE/dt \vert} = \frac{3 n k T}{2 n^2 \Lambda(T)} = \frac{3 k T}{2 n \Lambda(T)}
\end{equation}
where $R$ is the gas constant. A parcel of initially $10^6$ K gas interacting with an equal mass of $10^4$ K gas will, therefore, assuming thermal equilibrium is reached (so that $T_{\rm mixed} \approx 5 \times 10^5$ K and the density is twice the initial density of the halo gas), decrease its cooling time
by almost an order of magnitude. The cooling time for $10^6$ K gas at 0.1 solar metallicity is $\sim{}0.5$ Gyr \citep{SutherlandDopita1993}, so for the mixture $t_{\rm cool} \sim{} 100$ Myr.

The mixing of the disc gas with low-metallicity halo gas is also an important factor in matching observations. Studies of G-type, K-type, and M-type dwarf stars \citep{Schmidt1963, WortheyEtal1996, CasusoBeckman2004, WoolfWest2012} in the MW show that they possess a cumulative metallicity distribution that peaks strongly at solar metallicity, with a low-Z tail falling off much faster than the predictions of `closed-box' chemical evolution models, which find a Z-distribution that increases steadily with time and leads to a flatter slope. A solution to this discrepancy is near-continuous infall of metal-poor gas onto the Galactic disc, at a rate of $\sim 1 \msun$ yr$^{-1}$, in order to dilute the enrichment from stellar feedback in situ. The gas that is entrained in our model is of preferentially low metallicity, $0.1$ solar in the IC, compared to the disc gas metallicity of $1$ solar. As time goes on, some of this halo gas will increase in metallicity due to enrichment from SNe, but as can be seen in Figure 1 (bottom panel), low-Z gas from above and below the immediate mixing region replaces the original low-Z gas that has mixed. It is clear, therefore, that this mechanism can bring metal-poor gas into the Galaxy. The mixing of the solar metallicity gas with the metal-poor gas has an added effect of increasing the mixture's capacity to cool -- simply mixing temperatures will bring the halo gas closer to the peak of the cooling curve, but mixing metallicities will move this peak higher in terms of cooling rate \citep{KWH1996}.

With the high resolution that we employ, and with the use of a physical thermal conductivity, it is likely that in certain places we start to resolve the Field length \citep{Field1965, ArmillottaEtal2016} and see its effects on the evolution of the outflow. The Field length for cold gas structures embedded in a hot medium is given by,
\begin{equation}
\lambda_{\rm Field} \equiv \sqrt{\left(\frac{\kappa_{\rm Sp} T_{\rm hot}}{{n_{\rm cold}}^2 \Lambda(T_{\rm cold})}\right)}
\end{equation}
where $\kappa_{\rm Sp}$ is the Spitzer conductivity, $n_{\rm cold}$ and $T_{\rm cold}$ are the density and cooling rate of the cold gas respectively, and $T_{\rm hot}$ is the temperature of the hot ambient medium. For our initial parameters, with the fiducial 0.1 Spitzer conductivity, the Field length is $\sim 10$ pc. This quantity is important because it marks the boundary between a regime where thermal conduction dominates over radiative cooling, and vice versa.

For scales $l \ll \lambda_{\rm Field}$, the temperature evolution of the cold gas structures moving through the hot medium is dominated by thermal conduction, whereas for $l \gg \lambda_{\rm Field}$ it is the radiative cooling that dominates. The outflowing cold gas that interacts with the hot gas can transition between the two regimes as it moves through the hot medium and breaks up, with the evolution of larger cold gas structures in the outflow dominated by radiative cooling but the late-time behaviour of the smaller remnants of the outflow being more heavily influenced at their boundaries by the thermal conduction from the hot gas. In our simulations we can take the smoothing length as the spatial resolution element, which varies adaptively over the course of the run but is typically $\approx 10-20$ pc in the outflowing gas. In a few cases, gas within cold clumps at the edges of the outflow gets down to smoothing lengths of $\approx 2-5$ pc. In these cases, therefore, we can be reasonably confident that thermal conduction with the hot gas influences the evolution of the smaller cold gas structures. This means that at any point in time there are two possible outcomes; either the outflow gas survives and increases in mass as it entrains more and more of the hot medium, or it evaporates in the hotter medium. The inclusion of a physical thermal conductivity in our simulations is therefore a important feature, as is the reasonable convergence of the results with varying $\eta$ (see Figure 11). 

\subsection{A constant cold gas mass through positive feedback}\label{sec:flat}

The SFR of the MW is almost flat over its lifetime \citep[e.g.,][]{Rocha-PintoEtal2000}, which implies a roughly constant cold gas reservoir from which stars can form. Within the framework of our current understanding of galaxy formation, maintaining such a reservoir is a challenge after $z \sim 1$, since the Galaxy no longer receives significant amounts of additional cold gas from its wider cosmological environment. The gas that is used up in star formation must somehow be replenished, however, in order to keep the cold gas mass in the Galaxy approximately constant. As we have shown in Figure 2, with the positive feedback process extracting gas from the hot halo, we can maintain a roughly constant cold gas mass (blue line) for at least $3$ Gyr (the limit of the fiducial H1 run). Of course, since this is a simplified model, the full picture is likely to be more complex, but it provides strong evidence for positive feedback helping to keep this cold gas reservoir constant.

When we turn our attention to the \emph{extracted} gas, rather than the total (Figure 4), we notice that the cold gas extracted from the halo and the hot gas extracted from the disc (in the latter case, by simple SNe heating) have very different histories. The extracted cold gas is on the whole fairly monotonic, with the $\Mdot$ staying mostly in the positive range, with the occasional dip below zero. The hot gas on the other hand, oscillates largely around zero in the $\Mdot$, and ends up almost back where it started after 1 Gyr in the total mass (top plot). This makes sense when one considers the mechanism behind each mode -- gas cooled from the halo will remain cold until it can form stars and/or be heated by SNe feedback, which happens mostly in the disc. Gas heated from the disc by SNe, however, will cool relatively quickly since the gas is dense. As such we see peaks in the hot gas $\Mdot$ at around the same times as the peaks in the SFR (Figure 6, black line) when the starbursts occur, but these hot gas peaks are typically short lived.

The runs with different SF thresholds (see Table 1) show a similar form of the extracted cold gas mass and the corresponding time derivative as in our fiducial H1 run. In Figure 5 we plot each on the same axes, and we find that the extracted cold gas mass is largely the same until $\approx 250$ Myr, after which they start to diverge. The runs with an SF threshold above 0.1 atom per cc continue rising to 1 Gyr, while the runs with SF thresholds less than this flatten out (red, magenta, and green lines). In these latter two simulations, the SF threshold is so low that stars are able to form extremely easily, including in the cold gas that is sent out into the halo. Thus, the heating from SNe occurs not just each time the disc reforms but indeed across the whole duration of the simulation. This can be seen in Figure 6, where there is a clear difference between these two low SF threshold runs versus the others -- the red, magenta, and green lines are largely continuous on the SFR plot (bottom panel), while the high SF threshold runs (black, blue lines) are bursty, only spiking when the disc has reformed enough to undergo another starburst. This burstiness gives us the time interval for our galactic fountain cycle, which is $\approx 250-300$ Myr (for H1) Of course this differs for each run with different SFRs, since more powerful starbursts send the gas further out, causing it to take longer to return and re-form the disc.





\subsection{Positive feedback trend with SFR}\label{sec:trend}

\begin{figure*}
\centering
\includegraphics[width=1.0\textwidth]{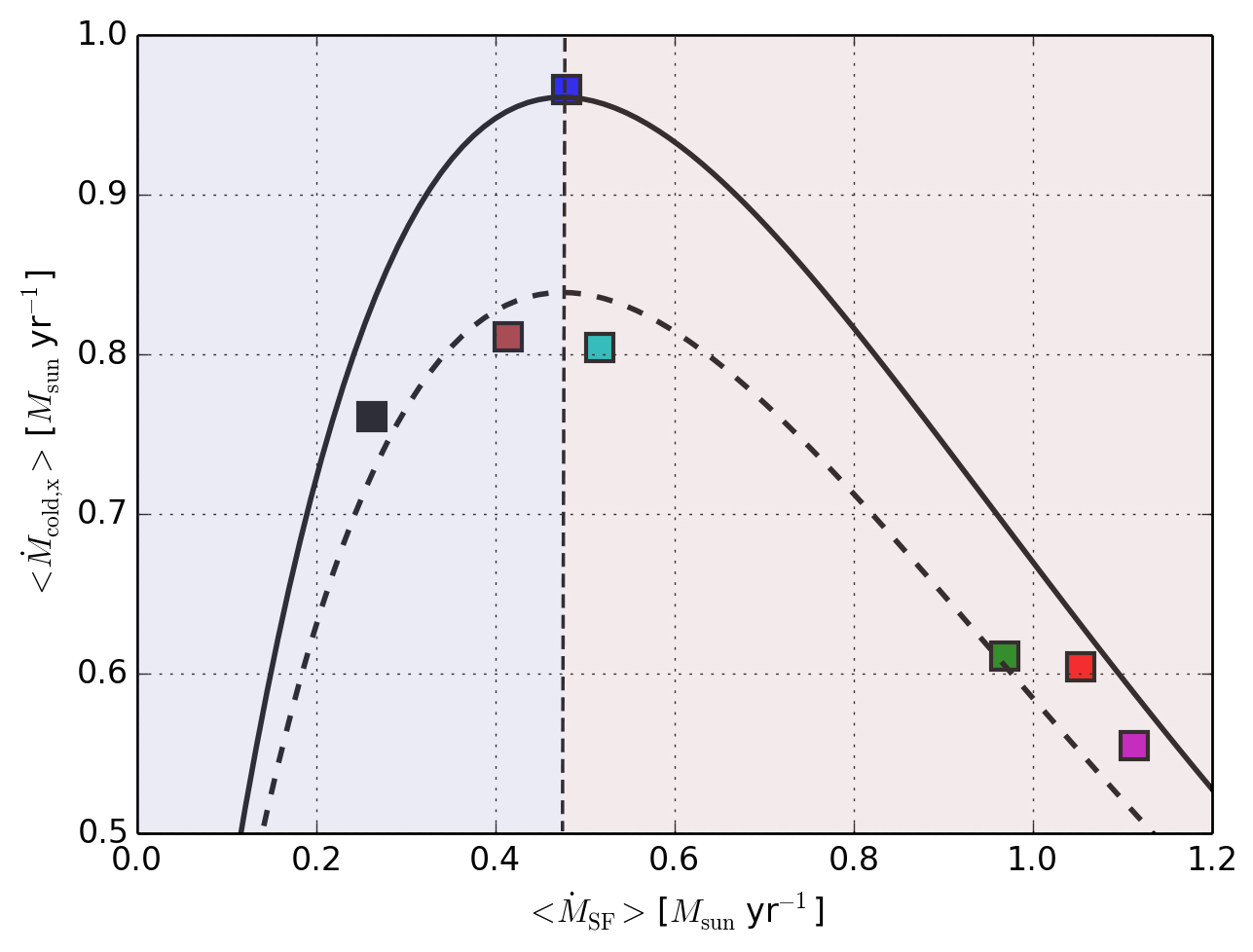}
\hspace{0.1in}
\begin{minipage}[b]{1.0\textwidth}
\caption[]{Plot showing the functional form of our simple analytic model described by Equation 9 compared with the averaged data points from the simulations. We show two normalisations of the analytic expression, with $A = 5.5$ (solid) and $A = 4.8$ (dashed). Qualitatively, the functional form agrees quite well with the data points (see Section 4.2).}
\end{minipage}
\label{fig:sftrendfit}
\end{figure*}

\begin{figure*}
\centering
\includegraphics[width=1.0\textwidth]{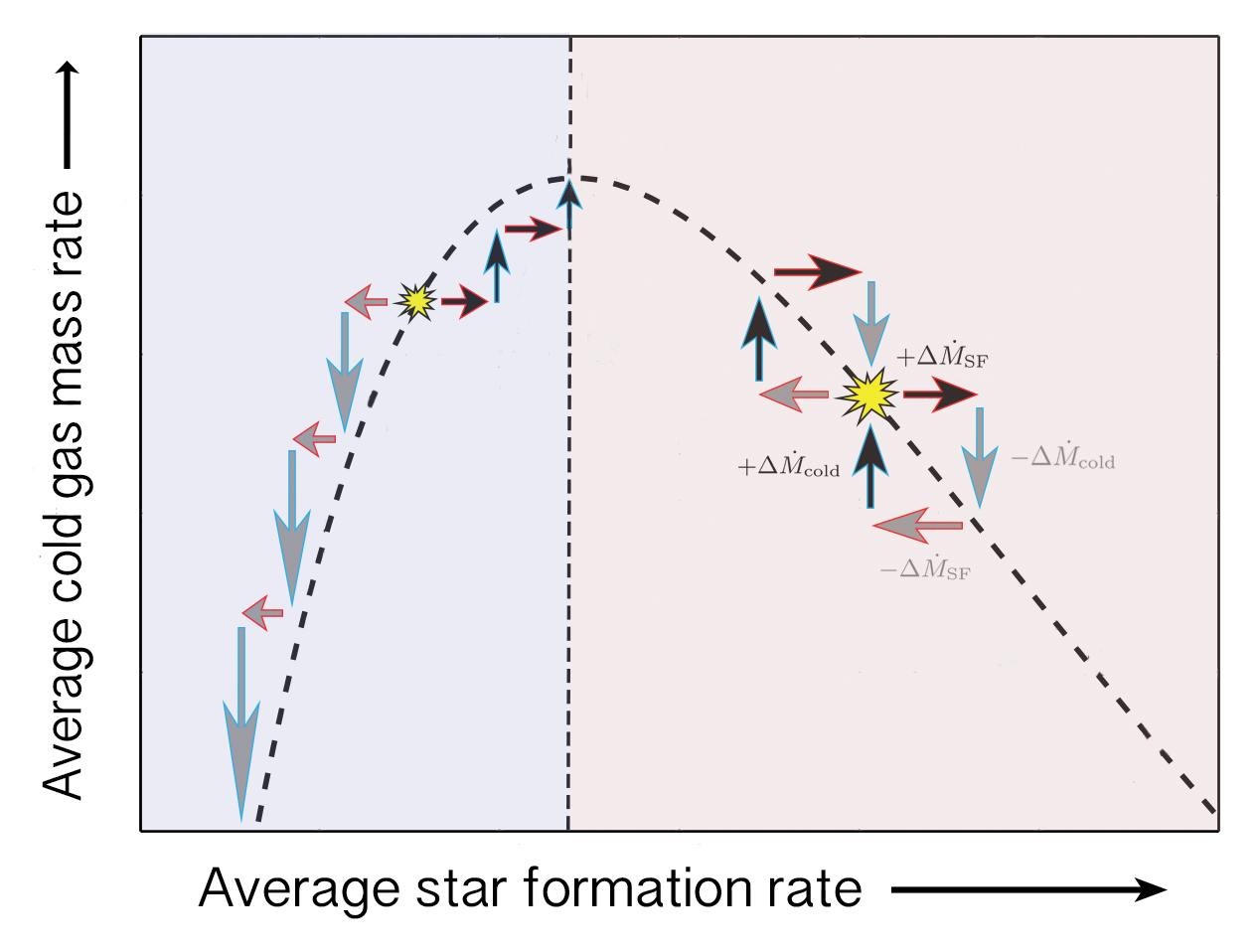}
\hspace{0.1in}
\begin{minipage}[b]{1.0\textwidth}
\caption[]{Cartoon demonstrating the stability of each side of the positive/negative feedback trend for a galaxy that lies on the trend line. Here we have used the analytical expression derived in Section 4.2, with a normalisation that is the average of the two normalisations shown in Figure 15. A galaxy that lies on the negative feedback trend (right-hand side) is very stable to perturbations in its average SFR. An increase (due perhaps to an external influence such as a merger or flyby) in average SFR would, assuming that the tendline is followed, lead to a decrease in the rate at which cold gas is extracted from the hot halo. This would in turn lead to a decrease in average SFR and a subsequent increase in the rate at which cold gas is extracted from the hot halo, bringing the galaxy back to where it started on the negative feedback trendline. Conversely, a perturbation that decreased the average SFR would lead to the same result but in the opposite direction. A galaxy that lies on the left-hand side of the plot, however (the positive feedback trend) would find any perturbation that either increased or decreased the SFR led to a continually increasing or decreasing extracted cold gas rate respectively. The position of the galaxy on this side of the trend would therefore be very unstable, tending to either increase in average SFR to the peak of the trend at $\approx 0.5 \msun$ or decrease to zero.}
\end{minipage}
\label{fig:cartoontrend}
\end{figure*}

In Table 2 we have recorded the averages of various mass rates over the first Gyr in units of solar masses per year for a number of the runs. The first thing to note about the numbers in this Table is due to the different definitions of `cold' and `hot' gas, the numbers for each run (along the row) should not be expected to add up to any constant value -- instead, a more instructive comparison is down each column, where we can see how each average rate (cold gas, hot gas, SF, extracted cold gas, extracted hot gas) changes as we change the threshold density for SF. The SF column, in particular, shows us that as the SF threshold density decreases, the average SFR increases -- this is not particularly surprising. The column highlighted in bold, however, which refers to the average mass rate of cold gas extracted from the halo gas, is more interesting in that the value increases in going from SF$_{\rm thresh} = 10$ to SF$_{\rm thresh} = 1$, at which it peaks, and starts to decline to the lower SF threshold values.

Earlier in the paper we defined a `positive feedback trend' to be increasing SFR $\rightarrow$ increasing extracted cold gas, and the `negative feedback trend' as vice versa. When we plot the average SFR vs. the average extracted cold gas rate (Figure 7), we see that we have two regimes -- a positive feedback trend at low SFR values and a negative feedback trend at high SFR values, with the peak somewhere close to $0.5 \msun$ yr$^{-1}$.\\

We can construct a simple analytical model to describe the functional form of this plot. From the simulations we have performed, we can identify two reservoirs of gas; cold (the star-forming disc gas) and hot (the ambient medium). To encapsulate the positive and negative effects of stellar feedback on the cooling of hot halo gas discussed earlier, we introduce two (not necessarily constant) coefficients denoting positive ($\epsilon_+$) and negative ($\epsilon_-$) feedback efficiency\footnote{Note that the latter is not the same as the `negative feedback' defined in Figure 4, which was the gas from the cold disc that was heated by the SNe. Here we are only dealing with gas extracted from the hot medium.}. Specifically, $\epsilon_-$ includes the preventive effect of heating the hot medium by SNe which counteracts the cold gas extraction from the hot halo via positive feedback.

Our basic assumption then is that the extraction rate of cold gas is given by the net efficiency $\epsilon=\epsilon_+ - \epsilon_-$ times the SFR, i.e., $\dot{M}_{\rm cold, x} = (\epsilon_+ - \epsilon_-) \, \dot{M}_{\rm SF}$. We can write this as
\begin{equation}
\dot{M}_{\rm cold, x} = \epsilon_+ \, f \dot{M}_{\rm SF},
\end{equation}
where $f \equiv (1 - \epsilon_-/\epsilon_+)$ is, in general, a function of SFR. If we examine this function $f$ we see that when $\epsilon_- = \epsilon_+$, the extraction of cold gas from the hot reservoir is completely opposed by the SNe reheating this gas, so that $f = 0$. We expect this scenario to occur at high SFRs. In contrast, when SFR are close to zero, we expect SNe heating to be sub-dominant compared to cold gas extraction, i.e. $\epsilon_- \ll \epsilon_+$. Hence, we expect $f\sim{}1$. A simple assumption is thus $f = e^{-\dot{M}_{\rm SF}/\tau}$. By taking the derivative and using the boundary between the two regimes as the stationary point (at $\dot{M}_{\rm SF} \approx 0.5$ -- see Figure 7), we see that $\tau \approx 0.5$.

Furthermore, we make the assumption that $\epsilon_+$ is approximately constant. $\epsilon_+$ is the slope of the trend at low SFR (since $f \approx 1$ at low SFR) and it measures the effectiveness of positive feedback for a given value of the SFR. Thus, a constant $\epsilon_+$ implies that we will get the same mass of extracted cold gas for a given outflow mass. As a first approximation, this appears reasonable.
Thus we have,



\begin{equation}
\dot{M}_{\rm cold, x} = A \, e^{-\dot{M}_{\rm SF}/\tau} \, \dot{M}_{\rm SF}
\end{equation}

\noindent where $A \equiv \epsilon_+$ is the normalisation constant. We show this functional form in Figure 15, alongside the averaged data points from the simulations, for two (arbitrary) normalisations for purposes of illustration. We can see that the functional form of our simple analytic model fits the data points reasonably well, although the data points do not all lie on the same normalisation. Given that we adopted one of many potential choices for the functional form of $f$, our toy model may be a useful guide to understand the behavior of our simulations qualitatively, but not quantitatively.

This result, although tentative due to the nature of our simplified model, has interesting implications for the evolution of late-type galaxies. If we consider the different SFRs occuring not in different galaxies but in the \emph{same} galaxy across its lifetime, then we can infer that while in the positive feeback trend regime (left part of the plot), any change to the SFR will be unstable -- an increase in SFR will lead to a greater amount of cold gas extracted from the hot halo, which will in turn lead to more star formation and a further increase in SFR. Vice versa, if the SFR drops for a period of time, less cold gas will be extracted, causing the SFR to drop still further. As a result it is likely that galaxies would not stay where they are in this left-hand part of the plot, but rather reach either peak cold gas extraction (at SFR $\approx 0.5 \msun$ yr $^{-1}$) or reach zero SFR (`red and dead'). These two values of the SFR are therefore `attractors' in the sense that any perturbation to the SFR would cause it to attain one of these values. We note, of course, that the values in this plot are averaged SFRs over 1 Gyr, rather than instantaneous SFRs, and so we cannot comment on how quickly a galaxy would move left or right along this slope.

In the right-hand part of the plot (negative feedback trend), however, the behaviour of the galaxy subject to any perturbation of the SFR is very stable -- 
an increase in SFR reduces the amount of new cold gas extracted from the halo, which subsequently causes the SFR to drop.
Likewise, a decrease in the SFR causes it to rise again. Thus galaxies can in theory remain where they are on the negative feedback trend for a long time. 

This simple stability discussion is depicted in cartoon schematic form in Figure 16. The peak of cold gas extraction rate that divides the two trends lies, in our suite of simulations, at $\approx 1 \msun$ yr$^{-1}$, which is in the approximate range for the required cold gas infall rate to solve the G-dwarf (and K- and M-dwarf) problem. The SFR for this peak is $\approx 1 \msun$ yr$^{-1}$, which is also in the ballpark range of the MW SFR over it's lifetime. However, we stress that, due to the nature of our somewhat limited model, we do not wish to make any quantitative conclusions in this paper -- it is entirely possible that even small modifications to our setup could result in the peak in the stability plot being shifted left or right in SFR and up or down in rate of cold gas extraction (not to mention that we currently lack error  bars on these measurements, which might alter these numbers also). For now we prefer to speculate on the qualitative form of this plot rather than the precise normalisation.

\subsection{Effectiveness of positive feedback}\label{sec:effect}

\begin{figure*}
\centering
\includegraphics[width=1.0\textwidth]{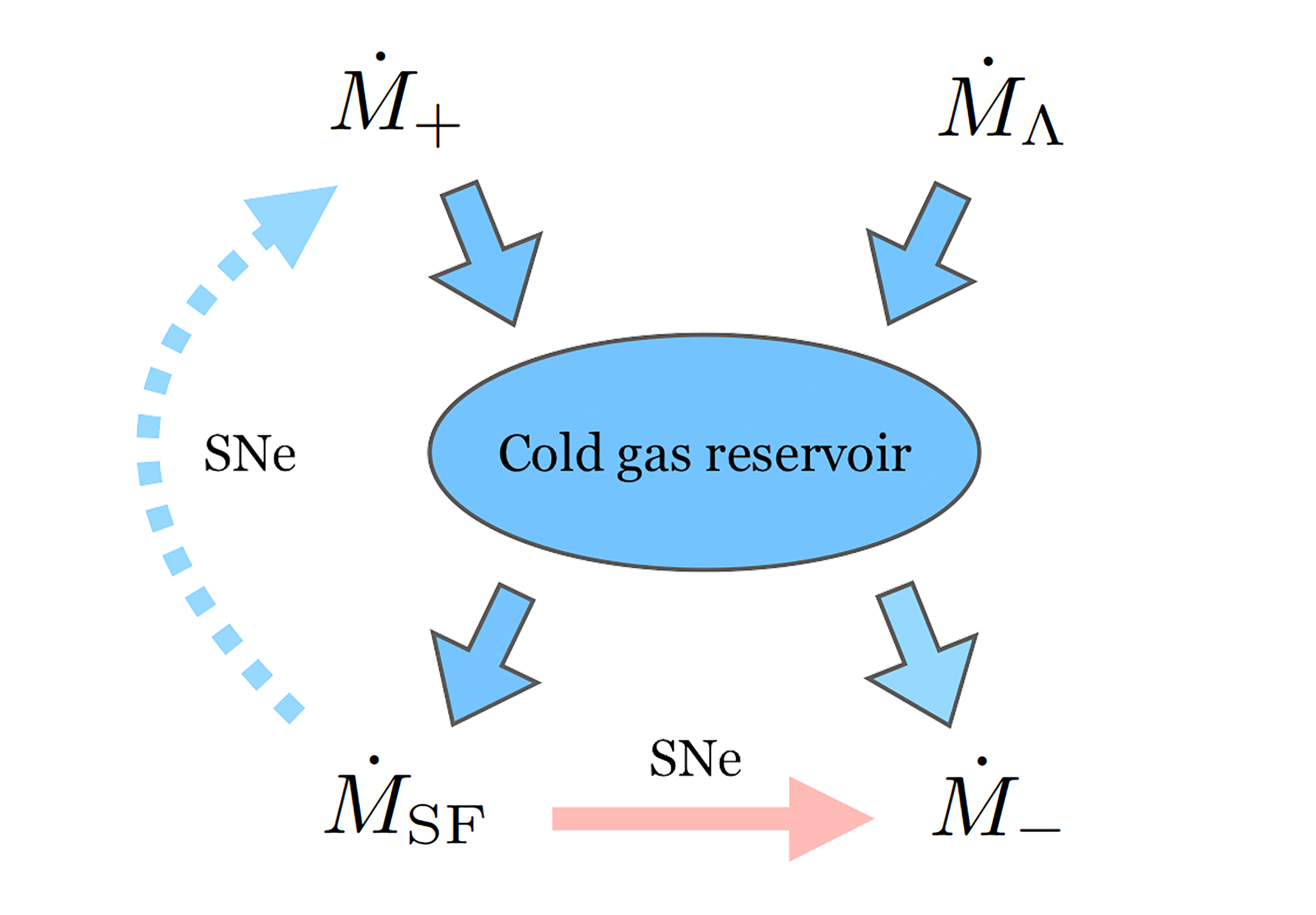}
\hspace{0.1in}
\begin{minipage}[b]{1.0\textwidth}
\caption[]{Cartoon schematic for a low-redshift (i.e., assuming no cosmological accretion) gas regulator model that includes positive feedback. The sources are radiative cooling ($\dot{M}_\Lambda$) and the positive feedback cold gas extraction mechanism outlined in this paper ($\dot{M}_{\rm +}$) and the sinks are star formation ($\dot{M}_{\rm SF}$) and negative feedback i.e., heating from SNe ($\dot{M}_{\rm -}$). Both the positive and negative feedback are caused by SF via the SNe, although in the case of the positive feedback this is more indirect. It should be noted that in this schematic we make no comment on the relative quantitative contribution of each source or sink.}
\end{minipage}
\label{fig:cartoon}
\end{figure*}

In \cite{ArmillottaEtal2016}, it was noted that the increase in cold gas mass as the cold cloud passed through the hot medium was strongly dependent on the temperature of the hot gas \citep[Figure 6 in][]{ArmillottaEtal2016}. These authors found that an increase of only a factor of 8 from the initial temperature of $10^6$ K reduced the amount of cold gas collected by the end of the 60 Myr period by a factor of $\sim 30$. To determine how a change in the ambient environment affects our positive feedback mechanism, we have plotted, against time, the ratio of the average temperature of the ambient medium at the approximate `mixing height' at time $t$ to the equivalent average temperature at that height at $t = 0$. This gives us the change in the average temperature of the hot gas (note: this is the `initially' hot gas that is tagged at the beginning of the simulation) that the cold gas must try to entrain as time goes on. This is shown in Figure 13 for each SF threshold. It is clear that the heating from the SNe, being confined to a relatively narrow periodic box, quickly heats up the ambient gas above the disc where the outflow tries to mix with it. The temperature stays relatively close to $\sim 10^6$ K for the first $200$ Myr, but subsequently it is heated to $\sim 10^7$ K and remains at around this temperature for the rest of the run.

Based on this finding, and the finding of \cite{ArmillottaEtal2016}, we find it logical to conclude that the reason for the reduction in rate of cold gas extraction after $t \approx 300-400$ Myr is a hot medium that is just too hot for this process to be effective. We can plot a measure of this effectiveness by plotting our temperature ratio (y-axis of Figure 13) against our cold gas extraction rate (y-axis of Figure 5) as a running average with time to obtain Figure 14, which shows us that when the temperature is a few times $10^6$ K, the outflow is effective at entraining cold gas, while at factors of $10-15$ times the initial $10^6$ K, this effectiveness drops substantially.

This is the reason that all of our runs show a substantial peak\footnote{note that since the rate plots are plotted with a log y-axis, this strong peak may not be immediately apparent} in the cooling rate of the extracted gas (see Figure 4 for an example) around $200$ Myr, corresponding to the initial outflow after the first starburst occurs in the ICs. Subsequently to this peak, the rate of extraction of cold gas drops by a factor of $\approx 5-10$ for the rest of the run. Given that this feature appears only within this initial period of time for every run, we explored the idea of treating this peak as an initial transient, that should be subtracted from our analysis, since perhaps the initial starburst is anomalous due to the setup of the ICs. However, upon further study we have concluded that this peak is not an effect of the initial, powerful starburst that should be subtracted, since as Figure 3 shows, the subsequent SFR peaks are roughly of equal magnitude to the initial one. \emph{This shows that the initial starburst is no more powerful than the ones that come afterwards as a result of the disc reforming each time and continuing the cycle.}

Instead, it is clear that our initial setup provides the `optimal conditions' for positive feedback to work. The gas hotter than a few $\times 10^6$ K is no longer able to cool down to sub $10^6$ K temperatures after mixing and thus retains both a high temperature and a long cooling time. Indeed, it is likely that the initial $300-400$ Myr is a more realistic model of the positive feedback process than the subsequent period for two reasons: (1) our relatively simple (and low) SF threshold prescription allows stars to form and explode as SNe in the outflow, contributing to the heating of the ambient gas -- this may not occur in real galaxies where the requirement for SF may be more stringent (2) the nature of the periodic boundaries in our setup causes the hot gas from the SNe to be `trapped' in the domain rather than escaping. Any process in real galaxies that reduces the amount of hot gas from SNe in the mixing region will contribute to a far more efficient mode of positive feedback. 




\subsection{An altered gas regulator model}


We attempt to construct a simple schematic of how positive feedback might be included in a gas regulator model \citep[e.g.][]{DaveEtal2012, LillyEtal2013, Feldmann2013, 2015MNRAS.449.3274F}. In the usual picture, the cold\footnote{although we note that most gas regulator models do not make a distinction on the temperature of the gas} gas reservoir is increased by cosmological accretion and decreased by SF and outflows. In our picture we assume that cosmological accretion has ended, and instead invoke the hot halo as the source of gas that can grow the reservoir. This can be achieved either through radiative cooling ($\dot{M}_{\Lambda}$) or through positive feedback ($\dot{M}_{\rm +}$). The sinks are the same as in other regulator models; gas being used up forming stars ($\dot{M}_{\rm SF}$) or ejected in outflows/negative feedback ($\dot{M}_{\rm -}$). This simple model is shown in Figure 17. The star formation now plays a dual role; through SNe feedback the SF can either give rise to negative feedback and help to deplete the reservoir, or it can give rise to positive feedback and help to grow the reservoir. Depending on the relative dominance of each mode, the reservoir will either decrease or increase respectively, which will lead to a similar decrease or increase in the SFR. This in turn will effect a change in the feedback (both positive and negative), with an increase in SFR giving rise to more SNe. As discussed in Section \ref{sec:trend}, and shown in Figure 16, when averaged over 1 Gyr timescales, while the average SFR is below $\approx 0.5 \msun$ yr$^{-1}$, increasing the reservoir is likely to lead to more positive feedback dominance (which then further increases the reservoir and the SFR). Thus the reservoir and the SFR grow or shrink relatively quickly in this regime. Above the $0.5 \msun$ yr$^{-1}$ value, however, negative feedback would likely start to dominate, but since it has a stabilising effect, the reservoir and SFR would reach approximately constant values.

This latter insight may have profound implications for galaxy evolution after $z \sim 1$. It is of course somewhat speculative, but if there indeed exists a slope SFR$-\dot{M}_{\rm cold}$ that is \emph{increasing} up to a particular value, it is likely that this value would be reached quickly, should the SFR begin to increase for whatever reason. Conversely, if the opposite occurred, it is not unfeasible that the SFR would quickly drop all the way to near-zero, as the amount of cold gas returned to the galaxy via this positive feedback process became less and less. A slope of decreasing SFR$-\dot{M}_{\rm cold}$ would, on the other hand, mark a phase of galaxy evolution that was roughly stable. This may have relevance to the standard `blue cloud -- green valley -- red sequence' picture of SF quenching in galaxy formation. 

The schematic shown in Figure 17 is of course merely one `sub-network' of the larger galaxy formation network -- it is the one relevant to our model, but the full picture would include contributions from, e.g., AGN feedback, cosmic rays, major and minor mergers, and of course cosmological accretion. 


\subsection{Caveats}\label{sec:caveats}

The simulations presented in this paper are simplified models, and many aspects of them are less than complete descriptions of the relevant physical processes. Here we list the relevant limitations of our model with respect to the real Universe.\\
\emph{Star formation.} Our SF prescription is only based on a density threshold and molecular gas fraction (see Section \ref{sec:method}). The formation of stars is therefore not dependant on whether the gas is Jeans unstable and as a result we may overestimate the number of stars formed. We stress however that our model is not intended to be a fully realistic description of the Galaxy, but rather a means to explore the effect of different star formation rates on the existence and effectiveness of positive feedback.\\
\emph{Boundary conditions.} The computational domain is restricted, using periodic boundaries in $x$ and $y$, which allow us to sample the domain without losing any gas through these boundaries, but the effect of this is likely to increase the confining pressure of the gas within the domain, leading to a greater amount of heating as the outflow is not able to expand as it would in a full galaxy setup. In $z$ we use an outflow boundary, which may introduce small numerical artifacts, although we have mitigated that by extending the vertical domain to $200$ kpc in each direction.\\
\emph{Initial conditions.} At the disc-ambient boundary of the ICs, we implemented a minor numerical fix that prevented mixing due to spurious pressure forces within the first 200 Myr. These pressure forces are a natural consequence of having two different particle/cell distributions next to each other, even despite ensuring that the hydrodynamic pressure was constant across the boundary. The effect of these spurious forces was to cause hot gas from the ambient medium to sink onto the disc, and so the fix reduced the acceleration onto the disc in the vertical direction by a factor of 100, for the gas between 0.2 and 0.4 kpc above the disc plane. After 200 Myr, during the first outflow phase of the galactic fountain cycle, the disc and ambient gas distributions had mixed sufficiently that the numerical correction was no longer needed. The effect of our correction is to actually reduce mixing at the disc-ambient boundary for the first 200 Myr, so we do not expect it to have any bearing on our conclusions.\\
\emph{Velocity profile.} We model the hot halo without a shear velocity that would account for the difference in rotational velocity between the disc and the halo. It is likely that including the correct rotational profile of the gas would serve to re-accrete the outflow gas faster onto the disc and change the periodicity of the Galactic fountain cycle in our model. We base this on work by \cite{FraternaliBinney2008}, who simulated galactic fountain clouds moving in the potentials of NGC 891 and NGC2403, where the modelled interaction of these clouds with the hot ambient gas was necessary in order to reproduce the vertical rotational lag. These authors found that the process of entraining gas from a hot halo with lower rotational velocity than the galactic disc caused the clouds to lose angular velocity and promoted their accretion onto the disc. 


\section{Conclusions}\label{sec:conclusions}

We have presented the results of a numerical model that seeks to understand the role of SNe outflows in increasing or decreasing the cold gas content of the MW (and by extension, other late-type spiral galaxies). We have employed a simplified setup of a 2 kpc $\times$ 2 kpc portion of the MW disc (present day) that is allowed to form stars which subsequently explode as SNe. From the series of numerical experiments we have conducted using this model we have found the following:
\begin{enumerate}
    \item Feedback from SNe can be both positive AND negative in the context of promoting or hindering SF in the Galaxy.
    \item The action of SNe feedback giving rise to an outflow of cold ($T < 2 \times 10^4$ K) gas causes additional cold gas to be collected from the originally hot ambient medium (presumed to be the hot halo gas surrounding the galaxy).
    \item SNe also heat up cold gas, reducing the supply, but their role in creating an outflow that cools and entrain the hot gas dominates over their role in heating cold gas.
    \item This process of positive feedback is strongly dependant on the temperature of the hot medium from which the gas is cooled and entrained, with $T \sim 10^6$ K allowing for an efficient positive feedback mechanism but $T \sim 10^7$ K reducing the effectiveness considerably.
    \item The ability of SNe outflows to increase the amount of cold gas in a galaxy may improve with increasing SFR, but only up to a point -- in our case, $0.5 \msun$ yr$^{-1}$ -- after which negative feedback dominates.
    \item The typical values for the rate of extraction of cold gas in our simulations ($0.5-1.0 \msun$ yr$^{-1}$) is in approximate agreement with the required rate for the MW to maintain its near-constant SFR since $z \sim 1$.
    \item The presence of a peak in the average SFR vs. average extracted cold gas plot, for which we have constructed a simple analytic model, has interesting implications (see Section \ref{sec:discussion}) for galaxy evolution after $z \sim 1$ when positive feedback may start to dominate the gas budget -- is this peak value an `attractor' that galaxies tend to reach as their SFR increases through the action of positive feedback? Is it just as likely that the SFR would decrease and lead to a rapid reduction of both positive feedback and SFR until the galaxy was `red and dead'? 
\end{enumerate}

Our results are a promising addition to the new but growing field of positive feedback in galaxy formation. We have drawn on the work of \cite{ArmillottaEtal2016}, and related papers, and demonstrated that the effect they observe in simulations of individual cold clouds moving through hot gas can also take place in the larger, galactic outflow context. Subsequently, we have demonstrated that such a positive feedback mechanism likely varies with average SFR in a way that places a galaxy undergoing this process in one of two evolution regimes. We plan to conduct future work in this area by (i) running high-resolution models of whole (isolated) galaxies and (ii) attempting to detect this process of positive feedback occuring in state-of-the-art cosmological zoom-in simulations such as those conducted by the FIRE collaboration \citep{FIREpaper}. 






\section{Acknowledgments}
The authors thank the referee for multiple suggestions that helped to improve the quality of the paper. AH thanks Pedro Capelo and Simon Lilly for useful insights, and Steffi Walch for the SILCC papers that inspired parts of this work. RF acknowledges financial support from the Swiss National Science Foundation (grant no 157591). This work was supported by a grant from the Swiss National Supercomputing Centre (CSCS) under project IDs s698, s797, s926 and uzh18.

\bibliographystyle{mnras}

\bibliography{references}

\begin{thebibliography}{}
\makeatletter
\relax
\def\mn@urlcharsother{\let\do\@makeother \do\$\do\&\do\#\do\^\do\_\do\%\do\~}
\def\mn@doi{\begingroup\mn@urlcharsother \@ifnextchar [ {\mn@doi@}
  {\mn@doi@[]}}
\def\mn@doi@[#1]#2{\def\@tempa{#1}\ifx\@tempa\@empty \href
  {http://dx.doi.org/#2} {doi:#2}\else \href {http://dx.doi.org/#2} {#1}\fi
  \endgroup}
\def\mn@eprint#1#2{\mn@eprint@#1:#2::\@nil}
\def\mn@eprint@arXiv#1{\href {http://arxiv.org/abs/#1} {{\tt arXiv:#1}}}
\def\mn@eprint@dblp#1{\href {http://dblp.uni-trier.de/rec/bibtex/#1.xml}
  {dblp:#1}}
\def\mn@eprint@#1:#2:#3:#4\@nil{\def\@tempa {#1}\def\@tempb {#2}\def\@tempc
  {#3}\ifx \@tempc \@empty \let \@tempc \@tempb \let \@tempb \@tempa \fi \ifx
  \@tempb \@empty \def\@tempb {arXiv}\fi \@ifundefined
  {mn@eprint@\@tempb}{\@tempb:\@tempc}{\expandafter \expandafter \csname
  mn@eprint@\@tempb\endcsname \expandafter{\@tempc}}}

\bibitem[\protect\citeauthoryear{{Anderson}, {Churazov}  \&
  {Bregman}}{{Anderson} et~al.}{2016}]{AndersonEtal2016}
{Anderson} M.~E.,  {Churazov} E.,   {Bregman} J.~N.,  2016, \mn@doi [\mnras]
  {10.1093/mnras/stv2314}, \href
  {http://adsabs.harvard.edu/abs/2016MNRAS.455..227A} {455, 227}

\bibitem[\protect\citeauthoryear{{Armillotta}, {Fraternali}  \&
  {Marinacci}}{{Armillotta} et~al.}{2016}]{ArmillottaEtal2016}
{Armillotta} L.,  {Fraternali} F.,   {Marinacci} F.,  2016, \mn@doi [\mnras]
  {10.1093/mnras/stw1930}, \href
  {http://adsabs.harvard.edu/abs/2016MNRAS.462.4157A} {462, 4157}

\bibitem[\protect\citeauthoryear{{Babyk}, {McNamara}, {Nulsen}, {Russell},
  {Vantyghem}, {Hogan}  \& {Pulido}}{{Babyk} et~al.}{2018}]{BabykEtal2018}
{Babyk} I.~V.,  {McNamara} B.~R.,  {Nulsen} P.~E.~J.,  {Russell} H.~R.,
  {Vantyghem} A.~N.,  {Hogan} M.~T.,   {Pulido} F.~A.,  2018, \mn@doi [\apj]
  {10.3847/1538-4357/aacce5}, \href
  {https://ui.adsabs.harvard.edu/abs/2018ApJ...862...39B} {862, 39}

\bibitem[\protect\citeauthoryear{{Bogd{\'a}n} et~al.,}{{Bogd{\'a}n}
  et~al.}{2013}]{BogdanEtal2013}
{Bogd{\'a}n} {\'A}.,  et~al., 2013, \mn@doi [\apj]
  {10.1088/0004-637X/772/2/97}, \href
  {http://adsabs.harvard.edu/abs/2013ApJ...772...97B} {772, 97}

\bibitem[\protect\citeauthoryear{{Bregman}}{{Bregman}}{1980}]{Bregman1980}
{Bregman} J.~N.,  1980, \mn@doi [\apj] {10.1086/157776}, \href
  {http://adsabs.harvard.edu/abs/1980ApJ...236..577B} {236, 577}

\bibitem[\protect\citeauthoryear{{Casuso} \& {Beckman}}{{Casuso} \&
  {Beckman}}{2004}]{CasusoBeckman2004}
{Casuso} E.,  {Beckman} J.~E.,  2004, \mn@doi [\aap]
  {10.1051/0004-6361:20034393}, \href
  {https://ui.adsabs.harvard.edu/abs/2004A&A...419..181C} {419, 181}

\bibitem[\protect\citeauthoryear{{Cavagnolo}, {Donahue}, {Voit}  \&
  {Sun}}{{Cavagnolo} et~al.}{2008}]{CavagnoloEtal2008}
{Cavagnolo} K.~W.,  {Donahue} M.,  {Voit} G.~M.,   {Sun} M.,  2008, \mn@doi
  [\apj] {10.1086/591665}, \href
  {https://ui.adsabs.harvard.edu/abs/2008ApJ...683L.107C} {683, L107}

\bibitem[\protect\citeauthoryear{{Cicone} et~al.,}{{Cicone}
  et~al.}{2018}]{CiconeEtal2018}
{Cicone} C.,  et~al., 2018, \mn@doi [\apj] {10.3847/1538-4357/aad32a}, \href
  {https://ui.adsabs.harvard.edu/abs/2018ApJ...863..143C} {863, 143}

\bibitem[\protect\citeauthoryear{{Colbrook}, {Ma}, {Hopkins}  \&
  {Squire}}{{Colbrook} et~al.}{2017}]{ColbrookEtal2017}
{Colbrook} M.~J.,  {Ma} X.,  {Hopkins} P.~F.,   {Squire} J.,  2017, \mn@doi
  [\mnras] {10.1093/mnras/stx261}, \href
  {http://adsabs.harvard.edu/abs/2017MNRAS.467.2421C} {467, 2421}

\bibitem[\protect\citeauthoryear{{Dav{\'e}}, {Finlator}  \&
  {Oppenheimer}}{{Dav{\'e}} et~al.}{2012}]{DaveEtal2012}
{Dav{\'e}} R.,  {Finlator} K.,   {Oppenheimer} B.~D.,  2012, \mn@doi [\mnras]
  {10.1111/j.1365-2966.2011.20148.x}, \href
  {https://ui.adsabs.harvard.edu/abs/2012MNRAS.421...98D} {421, 98}

\bibitem[\protect\citeauthoryear{{Dekel} \& {Mandelker}}{{Dekel} \&
  {Mandelker}}{2014}]{DekelMandelker2014}
{Dekel} A.,  {Mandelker} N.,  2014, \mn@doi [\mnras] {10.1093/mnras/stu1427},
  \href {https://ui.adsabs.harvard.edu/abs/2014MNRAS.444.2071D} {444, 2071}

\bibitem[\protect\citeauthoryear{{Dubinski}, {Narayan}  \&
  {Phillips}}{{Dubinski} et~al.}{1995}]{DubinskiNarayanPhillips1995}
{Dubinski} J.,  {Narayan} R.,   {Phillips} T.~G.,  1995, \mn@doi [\apj]
  {10.1086/175954}, \href
  {https://ui.adsabs.harvard.edu/abs/1995ApJ...448..226D} {448, 226}

\bibitem[\protect\citeauthoryear{{Feldmann}}{{Feldmann}}{2013}]{Feldmann2013}
{Feldmann} R.,  2013, \mn@doi [\mnras] {10.1093/mnras/stt851}, \href
  {https://ui.adsabs.harvard.edu/abs/2013MNRAS.433.1910F} {433, 1910}

\bibitem[\protect\citeauthoryear{{Feldmann}}{{Feldmann}}{2015}]{2015MNRAS.449.3274F}
{Feldmann} R.,  2015, \mn@doi [\mnras] {10.1093/mnras/stv552}, \href
  {https://ui.adsabs.harvard.edu/abs/2015MNRAS.449.3274F} {449, 3274}

\bibitem[\protect\citeauthoryear{{Fern{\'a}ndez}, {Joung}  \&
  {Putman}}{{Fern{\'a}ndez} et~al.}{2012}]{FernandezEtal2012}
{Fern{\'a}ndez} X.,  {Joung} M.~R.,   {Putman} M.~E.,  2012, \mn@doi [\apj]
  {10.1088/0004-637X/749/2/181}, \href
  {http://adsabs.harvard.edu/abs/2012ApJ...749..181F} {749, 181}

\bibitem[\protect\citeauthoryear{{Field}}{{Field}}{1965}]{Field1965}
{Field} G.~B.,  1965, \mn@doi [\apj] {10.1086/148317}, \href
  {http://adsabs.harvard.edu/abs/1965ApJ...142..531F} {142, 531}

\bibitem[\protect\citeauthoryear{{Fraternali} \& {Binney}}{{Fraternali} \&
  {Binney}}{2008}]{FraternaliBinney2008}
{Fraternali} F.,  {Binney} J.~J.,  2008, \mn@doi [\mnras]
  {10.1111/j.1365-2966.2008.13071.x}, \href
  {http://adsabs.harvard.edu/abs/2008MNRAS.386..935F} {386, 935}

\bibitem[\protect\citeauthoryear{{Fraternali} \& {Tomassetti}}{{Fraternali} \&
  {Tomassetti}}{2012}]{FraternaliTomassetti2012}
{Fraternali} F.,  {Tomassetti} M.,  2012, preprint, \href
  {http://adsabs.harvard.edu/abs/2012arXiv1207.0093F} {} (\mn@eprint {arXiv}
  {1207.0093})

\bibitem[\protect\citeauthoryear{{Gaensler}, {Madsen}, {Chatterjee}  \&
  {Mao}}{{Gaensler} et~al.}{2008}]{GaenslerEtal2008}
{Gaensler} B.~M.,  {Madsen} G.~J.,  {Chatterjee} S.,   {Mao} S.~A.,  2008,
  \mn@doi [] {10.1071/AS08004}, \href
  {http://adsabs.harvard.edu/abs/2008PASA...25..184G} {25, 184}

\bibitem[\protect\citeauthoryear{{Grcevich} \& {Putman}}{{Grcevich} \&
  {Putman}}{2009}]{GrcevichPutman2009}
{Grcevich} J.,  {Putman} M.~E.,  2009, \mn@doi [\apj]
  {10.1088/0004-637X/696/1/385}, \href
  {http://adsabs.harvard.edu/abs/2009ApJ...696..385G} {696, 385}

\bibitem[\protect\citeauthoryear{{Gronke} \& {Oh}}{{Gronke} \&
  {Oh}}{2018}]{GronkeOh2018}
{Gronke} M.,  {Oh} S.~P.,  2018, \mn@doi [\mnras] {10.1093/mnrasl/sly131},
  \href {https://ui.adsabs.harvard.edu/abs/2018MNRAS.480L.111G} {480, L111}

\bibitem[\protect\citeauthoryear{{Haardt} \& {Madau}}{{Haardt} \&
  {Madau}}{2012}]{HaardtMadau2012}
{Haardt} F.,  {Madau} P.,  2012, \mn@doi [\apj] {10.1088/0004-637X/746/2/125},
  \href {http://adsabs.harvard.edu/abs/2012ApJ...746..125H} {746, 125}

\bibitem[\protect\citeauthoryear{{Hippelein} et~al.,}{{Hippelein}
  et~al.}{2003}]{HippeleinEtal2003}
{Hippelein} H.,  et~al., 2003, \mn@doi [\aap] {10.1051/0004-6361:20021898},
  \href {http://adsabs.harvard.edu/abs/2003A%26A...402...65H} {402, 65}

\bibitem[\protect\citeauthoryear{{Hobbs}, {Read}, {Power}  \& {Cole}}{{Hobbs}
  et~al.}{2013}]{HobbsEtal2013}
{Hobbs} A.,  {Read} J.,  {Power} C.,   {Cole} D.,  2013, \mn@doi [\mnras]
  {10.1093/mnras/stt977}, \href
  {http://adsabs.harvard.edu/abs/2013MNRAS.434.1849H} {434, 1849}

\bibitem[\protect\citeauthoryear{{Hobbs}, {Read}  \& {Nicola}}{{Hobbs}
  et~al.}{2015}]{HobbsEtal2015}
{Hobbs} A.,  {Read} J.,   {Nicola} A.,  2015, \mn@doi [\mnras]
  {10.1093/mnras/stv1469}, \href
  {http://adsabs.harvard.edu/abs/2015MNRAS.452.3593H} {452, 3593}

\bibitem[\protect\citeauthoryear{{Hobbs}, {Read}, {Agertz}, {Iannuzzi}  \&
  {Power}}{{Hobbs} et~al.}{2016}]{HobbsEtal2016}
{Hobbs} A.,  {Read} J.~I.,  {Agertz} O.,  {Iannuzzi} F.,   {Power} C.,  2016,
  \mn@doi [\mnras] {10.1093/mnras/stw251}, \href
  {http://adsabs.harvard.edu/abs/2016MNRAS.458..468H} {458, 468}

\bibitem[\protect\citeauthoryear{{Hogan} et~al.,}{{Hogan}
  et~al.}{2017}]{HoganEtal2017b}
{Hogan} M.~T.,  et~al., 2017, \mn@doi [\apj] {10.3847/1538-4357/aa9af3}, \href
  {https://ui.adsabs.harvard.edu/abs/2017ApJ...851...66H} {851, 66}

\bibitem[\protect\citeauthoryear{{Hopkins}}{{Hopkins}}{2015}]{Hopkins2015}
{Hopkins} P.~F.,  2015, \mn@doi [\mnras] {10.1093/mnras/stv195}, \href
  {http://adsabs.harvard.edu/abs/2015MNRAS.450...53H} {450, 53}

\bibitem[\protect\citeauthoryear{{Hopkins}}{{Hopkins}}{2017a}]{HopkinsEtal2017a}
{Hopkins} P.~F.,  2017a, preprint, \href
  {http://adsabs.harvard.edu/abs/2017arXiv171201294H} {} (\mn@eprint {arXiv}
  {1712.01294})

\bibitem[\protect\citeauthoryear{{Hopkins}}{{Hopkins}}{2017b}]{Hopkins2017c}
{Hopkins} P.~F.,  2017b, arXiv e-prints, \href
  {https://ui.adsabs.harvard.edu/abs/2017arXiv171201294H} {p. arXiv:1712.01294}

\bibitem[\protect\citeauthoryear{{Hopkins}}{{Hopkins}}{2017c}]{Hopkins2017b}
{Hopkins} P.~F.,  2017c, \mn@doi [\mnras] {10.1093/mnras/stw3306}, \href
  {http://adsabs.harvard.edu/abs/2017MNRAS.466.3387H} {466, 3387}

\bibitem[\protect\citeauthoryear{{Hopkins}, {Kere{\v{s}}}, {O{\~n}orbe},
  {Faucher-Gigu{\`e}re}, {Quataert}, {Murray}  \& {Bullock}}{{Hopkins}
  et~al.}{2014}]{FIREpaper}
{Hopkins} P.~F.,  {Kere{\v{s}}} D.,  {O{\~n}orbe} J.,  {Faucher-Gigu{\`e}re}
  C.-A.,  {Quataert} E.,  {Murray} N.,   {Bullock} J.~S.,  2014, \mn@doi
  [\mnras] {10.1093/mnras/stu1738}, \href
  {https://ui.adsabs.harvard.edu/abs/2014MNRAS.445..581H} {445, 581}

\bibitem[\protect\citeauthoryear{{Hopkins} et~al.,}{{Hopkins}
  et~al.}{2018}]{HopkinsEtal2018}
{Hopkins} P.~F.,  et~al., 2018, \mn@doi [\mnras] {10.1093/mnras/sty674}, \href
  {http://adsabs.harvard.edu/abs/2018MNRAS.477.1578H} {477, 1578}

\bibitem[\protect\citeauthoryear{{Joung}, {Bryan}  \& {Putman}}{{Joung}
  et~al.}{2012}]{JoungEtal2012}
{Joung} M.~R.,  {Bryan} G.~L.,   {Putman} M.~E.,  2012, \mn@doi [\apj]
  {10.1088/0004-637X/745/2/148}, \href
  {http://adsabs.harvard.edu/abs/2012ApJ...745..148J} {745, 148}

\bibitem[\protect\citeauthoryear{{Kacprzak}, {Churchill}, {Steidel}  \&
  {Murphy}}{{Kacprzak} et~al.}{2008}]{KacprzakEtal2008}
{Kacprzak} G.~G.,  {Churchill} C.~W.,  {Steidel} C.~C.,   {Murphy} M.~T.,
  2008, \mn@doi [\aj] {10.1088/0004-6256/135/3/922}, \href
  {http://adsabs.harvard.edu/abs/2008AJ....135..922K} {135, 922}

\bibitem[\protect\citeauthoryear{{Katz}, {Weinberg}  \& {Hernquist}}{{Katz}
  et~al.}{1996}]{KWH1996}
{Katz} N.,  {Weinberg} D.~H.,   {Hernquist} L.,  1996, \mn@doi [\apjs]
  {10.1086/192305}, \href {http://adsabs.harvard.edu/abs/1996ApJS..105...19K}
  {105, 19}

\bibitem[\protect\citeauthoryear{{Kere{\v s}}, {Katz}, {Weinberg}  \&
  {Dav{\'e}}}{{Kere{\v s}} et~al.}{2005}]{KeresEtal2005}
{Kere{\v s}} D.,  {Katz} N.,  {Weinberg} D.~H.,   {Dav{\'e}} R.,  2005, \mn@doi
  [\mnras] {10.1111/j.1365-2966.2005.09451.x}, \href
  {http://adsabs.harvard.edu/abs/2005MNRAS.363....2K} {363, 2}

\bibitem[\protect\citeauthoryear{{Kim} et~al.,}{{Kim}
  et~al.}{2014}]{AGORApaper}
{Kim} J.-h.,  et~al., 2014, \mn@doi [\apjs] {10.1088/0067-0049/210/1/14}, \href
  {http://adsabs.harvard.edu/abs/2014ApJS..210...14K} {210, 14}

\bibitem[\protect\citeauthoryear{{Krumholz} \& {Gnedin}}{{Krumholz} \&
  {Gnedin}}{2011}]{KrumholzGnedin2011}
{Krumholz} M.~R.,  {Gnedin} N.~Y.,  2011, \mn@doi [\apj]
  {10.1088/0004-637X/729/1/36}, \href
  {http://adsabs.harvard.edu/abs/2011ApJ...729...36K} {729, 36}

\bibitem[\protect\citeauthoryear{{Leitner} \& {Kravtsov}}{{Leitner} \&
  {Kravtsov}}{2011}]{LeitnerKravtsov2011}
{Leitner} S.~N.,  {Kravtsov} A.~V.,  2011, \mn@doi [\apj]
  {10.1088/0004-637X/734/1/48}, \href
  {http://adsabs.harvard.edu/abs/2011ApJ...734...48L} {734, 48}

\bibitem[\protect\citeauthoryear{{Lilly}, {Le Fevre}, {Hammer}  \&
  {Crampton}}{{Lilly} et~al.}{1996}]{LillyEtal1996}
{Lilly} S.~J.,  {Le Fevre} O.,  {Hammer} F.,   {Crampton} D.,  1996, \mn@doi
  [\apjl] {10.1086/309975}, \href
  {http://adsabs.harvard.edu/abs/1996ApJ...460L...1L} {460, L1}

\bibitem[\protect\citeauthoryear{{Lilly}, {Carollo}, {Pipino}, {Renzini}  \&
  {Peng}}{{Lilly} et~al.}{2013}]{LillyEtal2013}
{Lilly} S.~J.,  {Carollo} C.~M.,  {Pipino} A.,  {Renzini} A.,   {Peng} Y.,
  2013, \mn@doi [\apj] {10.1088/0004-637X/772/2/119}, \href
  {https://ui.adsabs.harvard.edu/abs/2013ApJ...772..119L} {772, 119}

\bibitem[\protect\citeauthoryear{{Madau}, {Pozzetti}  \& {Dickinson}}{{Madau}
  et~al.}{1998}]{MadauEtal1998}
{Madau} P.,  {Pozzetti} L.,   {Dickinson} M.,  1998, \mn@doi [\apj]
  {10.1086/305523}, \href {http://adsabs.harvard.edu/abs/1998ApJ...498..106M}
  {498, 106}

\bibitem[\protect\citeauthoryear{{Magnani} \& {Smith}}{{Magnani} \&
  {Smith}}{2010}]{MagnaniSmith2010}
{Magnani} L.,  {Smith} A.~J.,  2010, \mn@doi [\apj]
  {10.1088/0004-637X/722/2/1685}, \href
  {http://adsabs.harvard.edu/abs/2010ApJ...722.1685M} {722, 1685}

\bibitem[\protect\citeauthoryear{{Marasco} \& {Fraternali}}{{Marasco} \&
  {Fraternali}}{2011}]{MarascoFraternali2011}
{Marasco} A.,  {Fraternali} F.,  2011, \mn@doi [\aap]
  {10.1051/0004-6361/201015508}, \href
  {http://adsabs.harvard.edu/abs/2011A\%26A...525A.134M} {525, A134}

\bibitem[\protect\citeauthoryear{{Marasco}, {Fraternali}  \&
  {Binney}}{{Marasco} et~al.}{2012}]{MarascoEtal2012}
{Marasco} A.,  {Fraternali} F.,   {Binney} J.~J.,  2012, \mn@doi [\mnras]
  {10.1111/j.1365-2966.2011.19771.x}, \href
  {http://adsabs.harvard.edu/abs/2012MNRAS.419.1107M} {419, 1107}

\bibitem[\protect\citeauthoryear{{Marinacci}, {Fraternali}, {Nipoti}, {Binney},
  {Ciotti}  \& {Londrillo}}{{Marinacci} et~al.}{2011}]{MarinacciEtal2011}
{Marinacci} F.,  {Fraternali} F.,  {Nipoti} C.,  {Binney} J.,  {Ciotti} L.,
  {Londrillo} P.,  2011, \mn@doi [\mnras] {10.1111/j.1365-2966.2011.18810.x},
  \href {http://adsabs.harvard.edu/abs/2011MNRAS.415.1534M} {415, 1534}

\bibitem[\protect\citeauthoryear{{Mastropietro}, {Moore}, {Mayer}, {Wadsley}
  \& {Stadel}}{{Mastropietro} et~al.}{2005}]{MastropietroEtal2005}
{Mastropietro} C.,  {Moore} B.,  {Mayer} L.,  {Wadsley} J.,   {Stadel} J.,
  2005, \mn@doi [\mnras] {10.1111/j.1365-2966.2005.09435.x}, \href
  {http://adsabs.harvard.edu/abs/2005MNRAS.363..509M} {363, 509}

\bibitem[\protect\citeauthoryear{{Melioli}, {Brighenti}, {D'Ercole}  \& {de
  Gouveia Dal Pino}}{{Melioli} et~al.}{2008}]{MelioliEtal2008a}
{Melioli} C.,  {Brighenti} F.,  {D'Ercole} A.,   {de Gouveia Dal Pino} E.~M.,
  2008, \mn@doi [\mnras] {10.1111/j.1365-2966.2008.13446.x}, \href
  {http://adsabs.harvard.edu/abs/2008MNRAS.388..573M} {388, 573}

\bibitem[\protect\citeauthoryear{{Miller} \& {Bregman}}{{Miller} \&
  {Bregman}}{2015}]{MillerBregman2015}
{Miller} M.~J.,  {Bregman} J.~N.,  2015, \mn@doi [\apj]
  {10.1088/0004-637X/800/1/14}, \href
  {http://adsabs.harvard.edu/abs/2015ApJ...800...14M} {800, 14}

\bibitem[\protect\citeauthoryear{{Nelson}, {Vogelsberger}, {Genel}, {Sijacki},
  {Kere{\v{s}}}, {Springel}  \& {Hernquist}}{{Nelson}
  et~al.}{2013}]{NelsonEtal2013}
{Nelson} D.,  {Vogelsberger} M.,  {Genel} S.,  {Sijacki} D.,  {Kere{\v{s}}} D.,
   {Springel} V.,   {Hernquist} L.,  2013, \mn@doi [\mnras]
  {10.1093/mnras/sts595}, \href
  {https://ui.adsabs.harvard.edu/abs/2013MNRAS.429.3353N} {429, 3353}

\bibitem[\protect\citeauthoryear{{Nicastro}, {Mathur}  \& {Elvis}}{{Nicastro}
  et~al.}{2008}]{NicastroEtal2008}
{Nicastro} F.,  {Mathur} S.,   {Elvis} M.,  2008, \mn@doi [Science]
  {10.1126/science.1151400}, \href
  {http://adsabs.harvard.edu/abs/2008Sci...319...55N} {319, 55}

\bibitem[\protect\citeauthoryear{{Puchwein} \& {Springel}}{{Puchwein} \&
  {Springel}}{2013}]{PuchweinSpringel2013}
{Puchwein} E.,  {Springel} V.,  2013, \mn@doi [\mnras] {10.1093/mnras/sts243},
  \href {https://ui.adsabs.harvard.edu/abs/2013MNRAS.428.2966P} {428, 2966}

\bibitem[\protect\citeauthoryear{{Putman}, {Peek}  \& {Joung}}{{Putman}
  et~al.}{2012}]{PutmanEtal2012}
{Putman} M.~E.,  {Peek} J.~E.~G.,   {Joung} M.~R.,  2012, \mn@doi [\araa]
  {10.1146/annurev-astro-081811-125612}, \href
  {http://adsabs.harvard.edu/abs/2012ARA\%26A..50..491P} {50, 491}

\bibitem[\protect\citeauthoryear{{Rafferty}, {McNamara}  \&
  {Nulsen}}{{Rafferty} et~al.}{2008}]{RaffertyEtal2008}
{Rafferty} D.~A.,  {McNamara} B.~R.,   {Nulsen} P.~E.~J.,  2008, \mn@doi [\apj]
  {10.1086/591240}, \href
  {https://ui.adsabs.harvard.edu/abs/2008ApJ...687..899R} {687, 899}

\bibitem[\protect\citeauthoryear{{Rocha-Pinto}, {Scalo}, {Maciel}  \&
  {Flynn}}{{Rocha-Pinto} et~al.}{2000}]{Rocha-PintoEtal2000}
{Rocha-Pinto} H.~J.,  {Scalo} J.,  {Maciel} W.~J.,   {Flynn} C.,  2000, \mn@doi
  [\apjl] {10.1086/312531}, \href
  {http://adsabs.harvard.edu/abs/2000ApJ...531L.115R} {531, L115}

\bibitem[\protect\citeauthoryear{{R{\"o}hser}, {Kerp}, {Lenz}  \&
  {Winkel}}{{R{\"o}hser} et~al.}{2016}]{RohserEtal2016}
{R{\"o}hser} T.,  {Kerp} J.,  {Lenz} D.,   {Winkel} B.,  2016, \mn@doi [\aap]
  {10.1051/0004-6361/201629141}, \href
  {http://adsabs.harvard.edu/abs/2016A\%26A...596A..94R} {596, A94}

\bibitem[\protect\citeauthoryear{{Sancisi}, {Fraternali}, {Oosterloo}  \& {van
  der Hulst}}{{Sancisi} et~al.}{2008}]{SancisiEtal2008}
{Sancisi} R.,  {Fraternali} F.,  {Oosterloo} T.,   {van der Hulst} T.,  2008,
  \mn@doi [\aapr] {10.1007/s00159-008-0010-0}, \href
  {http://adsabs.harvard.edu/abs/2008A\%26ARv..15..189S} {15, 189}

\bibitem[\protect\citeauthoryear{{Schmidt}}{{Schmidt}}{1963}]{Schmidt1963}
{Schmidt} M.,  1963, \mn@doi [\apj] {10.1086/147553}, 137, 758

\bibitem[\protect\citeauthoryear{{Shapiro} \& {Field}}{{Shapiro} \&
  {Field}}{1976}]{ShapiroField1976}
{Shapiro} P.~R.,  {Field} G.~B.,  1976, \mn@doi [\apj] {10.1086/154332}, \href
  {http://adsabs.harvard.edu/abs/1976ApJ...205..762S} {205, 762}

\bibitem[\protect\citeauthoryear{{Spitzer}}{{Spitzer}}{1942}]{Spitzer1942}
{Spitzer} Jr. L.,  1942, \mn@doi [\apj] {10.1086/144407}, \href
  {http://adsabs.harvard.edu/abs/1942ApJ....95..329S} {95, 329}

\bibitem[\protect\citeauthoryear{{Spitzer}}{{Spitzer}}{1962}]{Spitzer1962}
{Spitzer} L.,  1962, {Physics of Fully Ionized Gases}

\bibitem[\protect\citeauthoryear{{Stinson}, {Dalcanton}, {Quinn}, {Kaufmann}
  \& {Wadsley}}{{Stinson} et~al.}{2007}]{StinsonEtal2007}
{Stinson} G.~S.,  {Dalcanton} J.~J.,  {Quinn} T.,  {Kaufmann} T.,   {Wadsley}
  J.,  2007, \mn@doi [\apj] {10.1086/520504}, \href
  {https://ui.adsabs.harvard.edu/abs/2007ApJ...667..170S} {667, 170}

\bibitem[\protect\citeauthoryear{{Sutherland} \& {Dopita}}{{Sutherland} \&
  {Dopita}}{1993}]{SutherlandDopita1993}
{Sutherland} R.~S.,  {Dopita} M.~A.,  1993, \mn@doi [\apjs] {10.1086/191823},
  \href {https://ui.adsabs.harvard.edu/abs/1993ApJS...88..253S} {88, 253}

\bibitem[\protect\citeauthoryear{{Walch} et~al.,}{{Walch}
  et~al.}{2015}]{SILCCpaper}
{Walch} S.,  et~al., 2015, \mn@doi [\mnras] {10.1093/mnras/stv1975}, \href
  {http://adsabs.harvard.edu/abs/2015MNRAS.454..238W} {454, 238}

\bibitem[\protect\citeauthoryear{{Wilson}}{{Wilson}}{2013}]{Wilson2013}
{Wilson} C.~D.,  2013, in {Wong} T.,  {Ott} J.,  eds,  IAU Symposium Vol. 292,
  Molecular Gas, Dust, and Star Formation in Galaxies. pp 119--126,
  \mn@doi{10.1017/S174392131300077X}

\bibitem[\protect\citeauthoryear{{Woods}, {Wadsley}, {Couchman}, {Stinson}  \&
  {Shen}}{{Woods} et~al.}{2014}]{WoodsEtal2014}
{Woods} R.~M.,  {Wadsley} J.,  {Couchman} H.~M.~P.,  {Stinson} G.,   {Shen} S.,
   2014, \mn@doi [\mnras] {10.1093/mnras/stu895}, \href
  {https://ui.adsabs.harvard.edu/abs/2014MNRAS.442..732W} {442, 732}

\bibitem[\protect\citeauthoryear{{Woolf} \& {West}}{{Woolf} \&
  {West}}{2012}]{WoolfWest2012}
{Woolf} V.~M.,  {West} A.~A.,  2012, \mn@doi [\mnras]
  {10.1111/j.1365-2966.2012.20722.x}, \href
  {https://ui.adsabs.harvard.edu/abs/2012MNRAS.422.1489W} {422, 1489}

\bibitem[\protect\citeauthoryear{{Worthey}, {Dorman}  \& {Jones}}{{Worthey}
  et~al.}{1996}]{WortheyEtal1996}
{Worthey} G.,  {Dorman} B.,   {Jones} L.~A.,  1996, \mn@doi [\aj]
  {10.1086/118068}, \href
  {https://ui.adsabs.harvard.edu/abs/1996AJ....112..948W} {112, 948}

\makeatother
\end{thebibliography}

\end{document}